\documentclass[natbib, final]{CMDAarXiv}
\usepackage{a4wide}
\usepackage{amsmath, amssymb, bm}
\usepackage{amsfonts}
\usepackage{graphicx,float}
\usepackage{enumitem}
\usepackage{booktabs}
\usepackage[]{hyperref}

\newcommand{\phiarch}{\mathord{\buildrel{\lower3pt\hbox{$\scriptscriptstyle\frown$}}
\over \varphi }}

\journalname{Celestial Mechanics and Dynamical Astronomy}
 \title{Dynamics in first-order mean motion resonances: analytical study of a simple model with stochastic behaviour}
\titlerunning{Dynamics in first-order mean motion resonances}

 \author{S. Efimov \and
         V. Sidorenko}
 \institute{ S. Efimov
           \at Moscow Institute of Physics and Technology \\
					 \smallskip
                9 Institutskiy per., Dolgoprudny, Moscow Region, 141701, Russian Federation	\\
           \email{efimov.ss@phystech.edu}
           \and V. Sidorenko
           \at Keldysh Institute of Applied Mathematics\\
					     Russian Academy of Sciences, \\
					 \smallskip
               Miusskaya Sq., 4, 125047 Moscow, Russian Federation \\
						and \\	
					 \at Moscow Institute of Physics and Technology	\\
					 \smallskip
                9 Institutskiy per., Dolgoprudny, Moscow Region, 141701, Russian Federation}				
 \authorrunning{S.Efimov et al.}
 \date{}

\begin{document}

 \maketitle
 \begin{abstract}

We examine a 2DOF Hamiltonian system, which arises in study of first-order mean motion resonance in spatial circular restricted three-body problem ``star-planet-asteroid'', and point out some mechanisms of chaos generation.
Phase variables of the considered system are subdivided into fast and slow ones: one of the fast variables can be interpreted as resonant angle, while the slow variables are parameters characterizing the shape and orientation of the asteroid's orbit. Averaging over the fast motion is applied to obtain evolution equations which describe the long-term behavior of the slow variables. These equations allowed us to provide a comprehensive classification of the slow variables' evolution paths. The bifurcation diagram showing changes in the topological structure of the phase portraits is constructed and bifurcation values of Hamiltonian are calculated. Finally, we study properties of the chaos emerging in the system.


\end{abstract}

\section{Introduction}

The model system which will be considered below arises in studies of first-order mean motion resonances (MMR) in restricted three-body problem (R3BP) ``star-planet-asteroid''. If asteroid makes $p \in N$ revolutions around the star in the same amount of time in which the planet makes $p+1$ revolutions, there is an \emph{exterior} resonance of the first-order denoted as  $p:(p+1)$. The term exterior refers to the fact that in this case the asteroid's semi-major axis is larger than semi-major axis of the planet. The \emph{interior} MMR $(p+1):p$ takes place when asteroid makes $p+1$ revolutions during the time in which planet makes $p$. The first-order MMRs are quite common and, therefore, intensively studied by many specialists. The related bibliography is given in \cite{Gallardo2018} and \cite{Nesvorny2002}. In particular, much effort has been spent to reveal why $2:1$ resonance with Jupiter corresponds to one of the largest gaps in the main asteroid belt (so-called Hecuba gaps), whereas $3:2$ MMR resonance is populated by numerous objects of Hilda group, and it is also very likely, that Thule group of objects in $4:3$ MMR is rather large \citep{Broz2008,Henrard1996,Lemaitre1990}. The discoveries of trans-Neptunian objects made it urgent to study the exterior resonances with Neptune: \emph{twotino} (MMR $1:2$) and \emph{plutino} (MMR $2:3$) form big subpopulations in the Kuiper belt \citep{Li2014a, Li2014b, Nesvorny2000, Nesvorny2001}.

It is possible to construct a model of dynamics in first-order MMR, taking into account only the leading terms in the Fourier series expansion of disturbing function \citep{FerrazMello84, Wisdom86, Gerasimov1990}. However, studies of planar R3BP \citep{Beauge94, Jancart02} revealed, that some important characteristics of first-order MMRs are reproduced only when the second-order Fourier terms are accounted for. In this paper we concentrate our attention on that part of a phase space, where eccentricities and inclinations are in relation $e\ll i \ll1$ (this, in some sense, is a case opposite to the planar problem, for which the relation is $e>i=0$). We intend to demonstrate, that in non-planar case second-order terms are no less important, as they make a model essentially stochastic. In contrast, the first-order models are proven to be integrable \citep{FerrazMello84}, and thus cannot reproduce chaotic dynamics found in multiple numerical studies of first-order resonances \citep{Wisdom88, Giffen73, Winter97a, Winter97b, Wisdom87},

There are different mechanisms for generating chaos in the dynamics of celestial bodies \citep{Holmes1990,Lissauer1999,Morbidelli}. Presence of MMR may lead to the so-called adiabatic chaos \citep{Wisdom85}, which is caused, roughly speaking, by small quasi-random jumps between regular phase trajectories in certain parts of the phase space, where adiabatic approximation is violated. Applying systematically Wisdom's ideas to study of MMRs \citep{Sidorenko_31, Sidorenko_QS, Sidorenko_JT}, we found that adiabatic chaos often coexists with the quasi-probabilistic transitions between specific phase regions. Both phenomena occur in that part of the phase space, where the ``pendulum'' or first-order Second Fundamental Model for Resonance \citep{Henrard83} approximations fail, because the first harmonic in the disturbing function Fourier series is not dominant. The goal of this paper is to carry out a comprehensive analysis of the introduced second-order model and investigate described mechanisms of chaotization, which, in our opinion, have not received proper attention in the past.

The paper is organized as follows. In Section \ref{Sec:modelH} the model Hamiltonian system, which has a structure of slow-fast system, is introduced. In Section \ref{Sec:fast} the fast subsystem is studied. Equations of motion for slow subsystem are constructed in Section \ref{Sec:slow} and their solutions are analyzed in Section \ref{Sec:PP}. Section \ref{Sec:chaos} is devoted to different chaotic effects present in the discussed model. In Section \ref{Sec:num} numerical evidence for existence of described phenomena is shown. The results are summarized in the last section. In Appendix A we reveal how the proposed model was derived. Details of the averaging procedure are elucidated in Appendix B.

\section{Model Hamiltonian system}\label{Sec:modelH}

We are dealing with 2DOF Hamiltonian systems with specific symplectic structure:
\begin{equation}\label{(1)}
\begin{aligned}
&\frac{{d\varphi }}{{d\tau}} = \frac{{\partial \Xi }}{{\partial \Phi }},
~
&\frac{{d\Phi }}{{d\tau}} =  - \frac{{\partial \Xi }}{{\partial \varphi }},
\\
&\frac{{dx}}{{d\tau}} = \varepsilon \frac{{\partial \Xi }}{{\partial y}},
~
&\frac{{dy}}{{d\tau}} =  - \varepsilon \frac{{\partial \Xi }}{{\partial x}}.
\end{aligned}
\end{equation}
The Hamiltonian $\Xi$ in \eqref{(1)} is expressed by
\begin{equation}\label{(2)}
\Xi (x,y,\varphi ,\Phi ) = \frac{{{\Phi ^2}}}{2} + W(x,y,\varphi ),~
W(x,y,\varphi ) = x\cos \varphi  + y\sin \varphi  + \cos 2\varphi.
\end{equation}
Appendix A describes in detail how the system \eqref{(1)}-\eqref{(2)} arises in studies of first-order MMR in three-dimensional R3BP. Here we only note that
\[
\varepsilon\sim{\mu ^{1/2}},\quad x\propto e\cos \omega,\quad y\propto -e\sin \omega,
\]
where $\mu\ll1$ is the fraction of the planet's mass in the total mass of the system, $e$~ and~$\omega$ denote the eccentricity and the argument of pericenter of asteroid's osculating orbit respectively. Further $\varepsilon$ is treated as a small parameter of the problem.
	Since in general variables $\varphi, \Phi, x, y$ vary with different rates ($d\varphi /d\tau, d\Phi /d\tau\sim 1$, while $dx/d\tau, dy/d\tau\sim\varepsilon\ll1$), we can distinguish in \eqref{(1)} \emph{fast} subsystem (described by the first line of equations) and \emph{slow} subsystem (the second line).

In limiting case $\varepsilon=0$ equations of fast subsystem coincide with equations of motion for particle with unit mass in a field with potential $W(x,y,\varphi)$, where $x$ and $y$ are treated as parameters. Let
\begin{equation}\label{(3)}
\varphi (\tau,x,y,\xi ),\quad \Phi (\tau,x,y,\xi )
\end{equation}
be a solution of fast subsystem with fixed values of $x$, $y$, and value $\xi$ of Hamiltonian~$\Xi$, which is the first integral of system \eqref{(1)}. In general the resonant angle $\varphi$ in \eqref{(3)} can \emph{librate} between two constant values or change monotonously through the whole interval $\left[0,2 \pi\right)$, i.e. \emph{circulate}. In either case
\[
\varphi (\tau + T,x,y,\xi ) = \varphi (\tau,x,y,\xi )\bmod \left(2\pi\right),
\]
where $T(x,y,\xi )$ is a period of the solution \eqref{(3)}.

Because fast variables vary much faster then the slow ones, the right-hand sides in differential equations of slow subsystem in \eqref{(1)} can be replaced by their average values along the solution~\eqref{(3)}. This yields the \emph{evolution equations}, which describe secular variations of $x$ and $y$ in closed form:
\begin{equation}\label{(4)}
\frac{{dx}}{{d\tau}} = \varepsilon \left\langle {\frac{{\partial \Xi }}{{\partial y}}} \right\rangle ,~\frac{{dy}}{{d\tau}} =  - \varepsilon \left\langle {\frac{{\partial \Xi }}{{\partial x}}} \right\rangle .
\end{equation}
Here
\begin{equation}\label{(5)}
\left\langle \Lambda \right\rangle  = \frac{1}{{T(x,y,\xi )}}\int\limits_0^{T(x,y,\xi )} {\Lambda (x,y,\varphi (\tau,x,y,\xi ))d\tau}.
\end{equation}

The solution \eqref{(3)} has an action integral:
\begin{equation}\label{(6)}
J(x,y,\xi ) = \frac{1}{{2\pi }}\int\limits_0^{T(x,y,\xi )} {{\Phi ^2}(\tau,x,y,\xi )d\tau}.
\end{equation}
For $\varepsilon\ne0$ function $J(x,y,\xi ) $ becomes an adiabatic invariant of slow-fast system~\eqref{(1)}. For a fixed $\xi$ trajectories of averaged equations \eqref{(4)} on a phase plane $(x,y)$ go along the lines with constant values of $J$. This allows to classify evolution equations \eqref{(4)} as adiabatic approximation \citep{Neishtadt_kirkwood, Wisdom85}.

In the next Sections we go through all the steps in construction of evolution equations via described approach and analyze in detail the behaviour of slow variables on different levels of Hamiltonian $\Xi=\xi$.

\section{Properties of fast subsystem's solutions for different values of slow variables (limiting case $\varepsilon=0$)}\label{Sec:fast}

\subsection{Partition of the plane $(x,y)$ based on the number of librating solutions}

Because the variables $x$ and $y$ change very slowly \eqref{(1)}, they can be treated as constant parameters, when considering the motion of the fast subsystem. Then the potential $W$ is just a ``two-harmonic'' function of $\varphi$ defined on circle $S^1$, and the motion in such potential can be described in terms of elliptic functions.

\begin{figure}[ht]
\includegraphics[width=122mm]{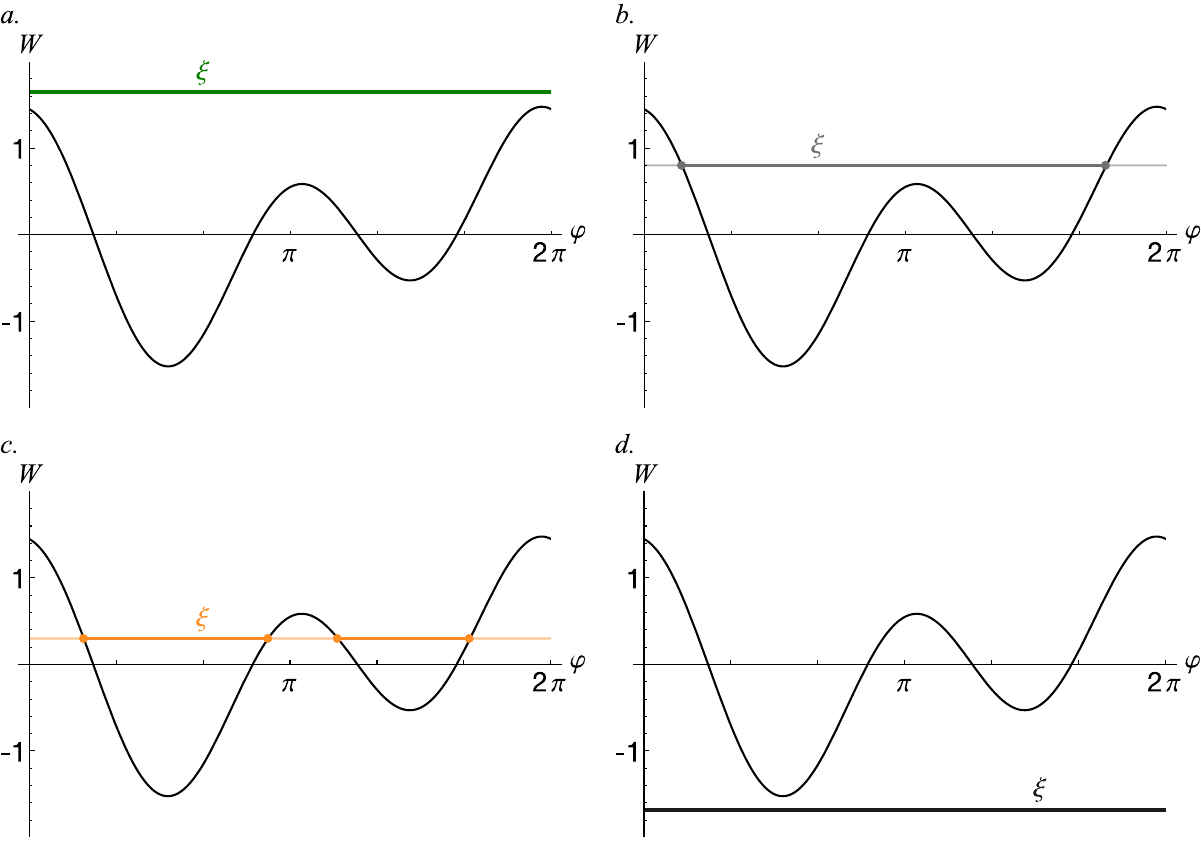}
\caption{Levels $\xi$ of Hamiltonian $\Xi$ corresponding to different types of fast subsystem's motion:
a. circulation, b. libration, c. two coexisting librating solutions, d. the motion is impossible}
\label{fig1}
\end{figure}

There are different types of motion depending on the Hamiltonian level $\xi$ at which it occurs (Fig.~\ref{fig1}). For us it is important, that for some values of  $x$ and $y$ two different librating solutions can exist on the same $\xi$ level (Fig.~\ref{fig1}c). This situation can take place when $W(\varphi)$ has four extrema on $S^1$.

A necessary condition for extremum ${{\partial W} {\left/
 {\vphantom {{\partial W} {\partial \varphi  = 0}}} \right.
 \kern-\nulldelimiterspace} {\partial \varphi  = 0}}$ after the replacement $\lambda=\tan \left(\varphi /2 \right)$ yields
\begin{equation}\label{(7)}
y{\lambda ^4} + 2(x + 4){\lambda ^3} + 2(x - 4)\lambda  - y = 0.
\end{equation}
Let ${\cal A}$ denote a region on the plane $(x,y)$, in which $W(\varphi)$ has four extrema. The equation \eqref{(7)} has four real roots inside this region and only two outside. Thus on the border of the region ${\cal A}$ the number of unique real roots is $3$ (with the exception of finite number of points in which there is only one unique real root) and the discriminant of~\eqref{(7)} is equal to zero. Therefore the equation for the border of the region ${\cal A}$ is
\begin{equation}\label{(8)}
{x^6} + 3{x^4}{y^2} - 48{x^4} + 3{x^2}{y^4} + 336{x^2}{y^2} + 768{x^2} + {y^6} - 48{y^4} + 768{y^2} - 4096 = 0.
\end{equation}
\begin{figure}[ht]
\center{
\includegraphics[width=122mm]{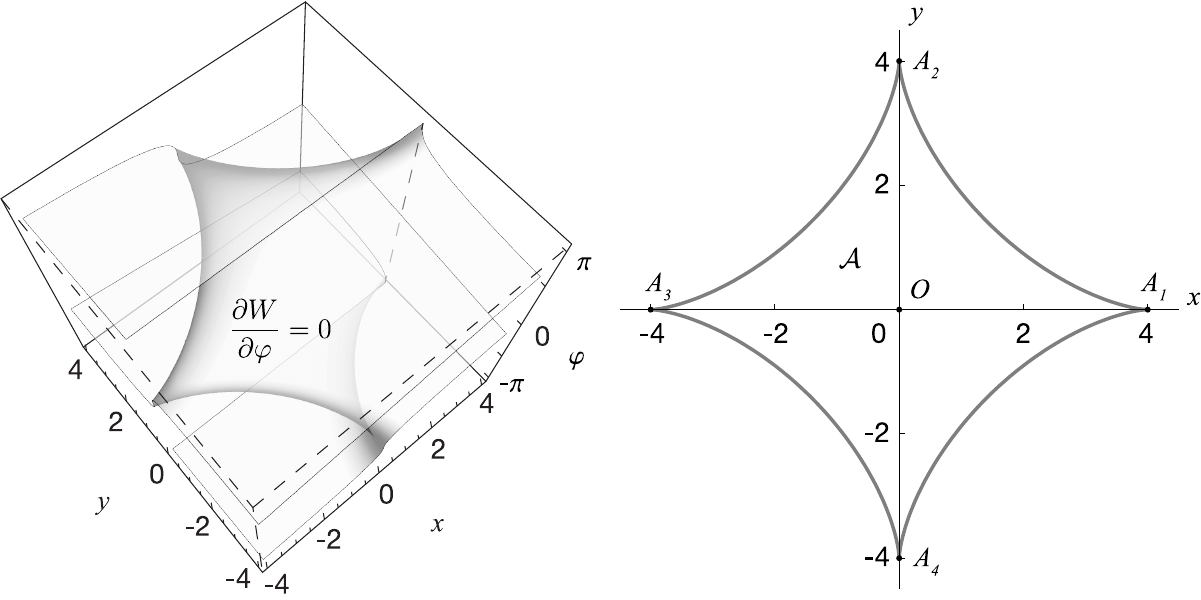}}
\caption{Extremal surface of potential $W(x,y,\varphi)$ (left), and astroid bounding the region ${\cal A}$, where $W$ has four extrema as a function of $\varphi$ on $S^1$ (right). $A_1,\ldots  , A_4$ -- astroid's cusps}
\label{fig2}
\end{figure}
By collecting the parts of this equation into perfect cube \eqref{(8)} is transformed to canonical algebraic equation of astroid (Fig.~\ref{fig2}):
\begin{equation}\label{(9)}
{\left( {{x^2} + {y^2} - {4^2}} \right)^3} + 27 \cdot {4^2}{x^2}{y^2} = 0.
\end{equation}
Which can be further reduced to
\begin{equation}\label{astroid}
{x^{2/3}} + {y^{2/3}} = {4^{2/3}}.
\end{equation}
It is convenient to use this astroid for the reference on the phase plane $(x,y)$.

\subsection{Critical curve partitioning the plane $(x,y)$ into regions with different types of fast subsystem' motion}
Let us introduce the notations ${W_{min}}(x,y)$ and ${W_{max}}(x,y)$ for global minimum and maximum of function $W$ on $S^1$ for given values of slow variables. If $(x,y) \in {\cal A}$, then $W$ has the second pair of minimum and maximum, which we shall denote ${W^*_{min}}(x,y)$ and ${W^*_{max}}(x,y)$ respectively.
Using these auxiliary functions, we can partition the $(x,y)$ plane for a given $\xi$ into different regions based on the type of fast subsystem's motion:
\[
\begin{aligned}
&Q_0=\left\{(x,y)~\left|~\xi<W_{min}\right.\right\},
\\
&Q_1=\left\{(x,y)\notin{\cal A}~\left|~
\xi\in\left(W_{min},\vphantom{W^*_{min}}W_{max}\right)\right.\right\}\bigcup
\\
&\hspace{.8cm}\left\{(x,y)\in{\cal A}~\left|~
\xi\in\left(W_{min},\vphantom{W^*_{min}}W_{max}\right)\backslash \left(W^*_{min},W^*_{max}\right)\right.\right\},
\\
&Q_2=\left\{(x,y)\in{\cal A}~\left|~
\xi\in\left(W^*_{min},W^*_{max}\right)\right.\right\},
\\
&Q_3=\left\{(x,y)~\left|~\xi>W_{max}\right.\right\}.
\end{aligned}
\]
The region ${Q_0}(\xi )$ will be called a \emph{forbidden region}, because inside of it $\Xi<\xi$ for any values of fast variables and fast subsystem has no solutions (Fig.~\ref{fig1}d). Region ${Q_1}(\xi )$ is the region with the single librating solution (Fig.~\ref{fig1}b), ${Q_2}(\xi )$ is the region with two librating solutions at given level $\xi$ (Fig.~\ref{fig1}c), and finally the region ${Q_3}(\xi)$ is where fast subsystem's solution circulates (Fig.~\ref{fig1}a). Illustrations for regions $Q_0(\xi), \ldots, Q_3(\xi)$ will follow.
\begin{figure}[ht]
\includegraphics[width=122mm]{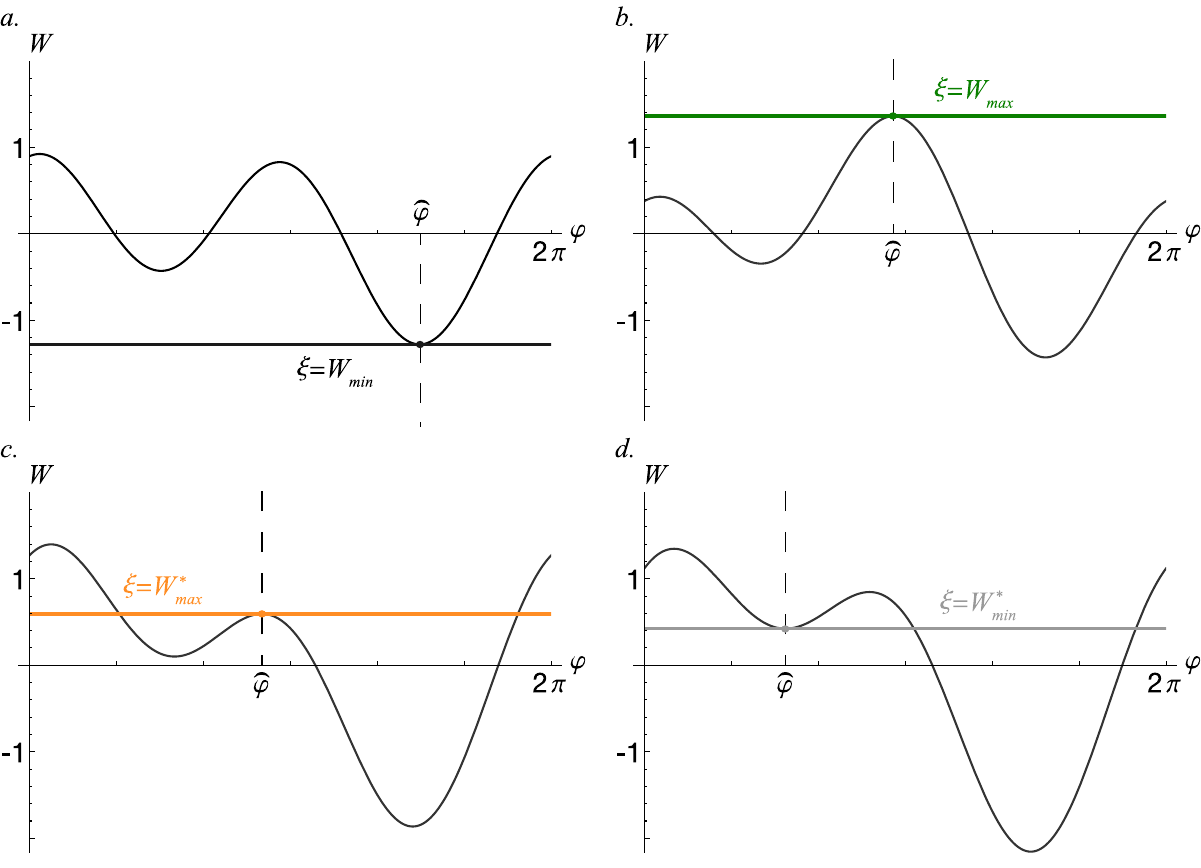}
\caption{Tangency of Hamiltonian level $\xi$ and different extrema of $W(\varphi)$, which occurs on the borders of regions $Q_0(\xi), \ldots, Q_3(\xi)$}
\label{fig3}
\end{figure}

\begin{figure}[H]
\center{\includegraphics[width=122mm]{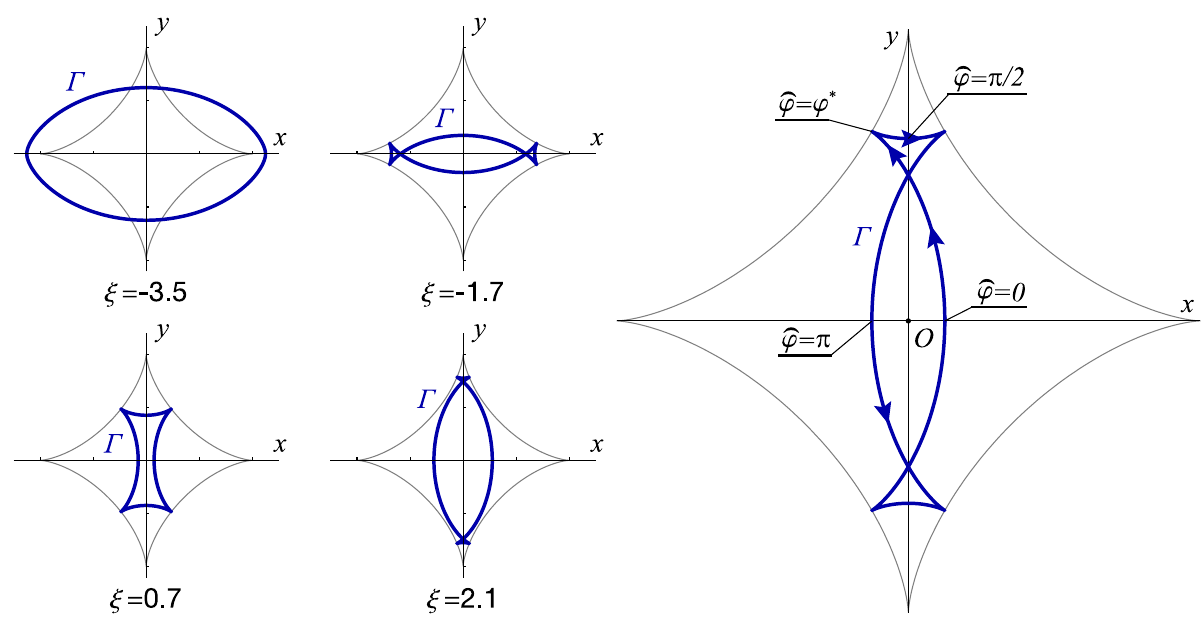}}
\caption{Critical curve $\Gamma(\xi)$: shape of the curve for different $\xi$ values (left), and parametrization of the curve by $\phiarch$ with arrows showing the direction in which the parameter increases (right)}
\label{fig45}
\end{figure}

Before that let us consider a border $\Gamma(\xi)$ between these regions. In every point of the border value of $W$ is equal to $\xi$ in one of its critical points (cf. Figures~\ref{fig1} and \ref{fig3}), which is why we shall call $\Gamma(\xi)$ a \emph{critical curve}.
After replacement $\lambda=\tan \left(\varphi /2 \right)$ the equation $W(\varphi)= \xi$ transforms into algebraic equation:
\begin{equation}\label{(10)}
(1 - \xi  - x){\lambda ^4} + 2y{\lambda ^3} - 2(\xi  + 3){\lambda ^2} + 2y\lambda  + (x + 1 - \xi ) = 0.
\end{equation}
As Figure~\ref{fig3} demonstrates, the point $(x,y)$ lies on the critical curve when equation \eqref{(10)} have at least one multiple real root, which is equivalent to discriminant of \eqref{(10)} being equal to zero:
\begin{multline}\label{(11)}
D(x,y,\xi ) = 64{\xi ^4} - 128{\xi ^2} - {x^6} + {\xi ^2}{x^4} - 18\xi {x^4} - 3{x^4}{y^2} - 15{x^4} + \\
 + 16{\xi ^3}{x^2} - 80{\xi ^2}{x^2} - 144\xi {x^2} - 3{x^2}{y^4} + 2{\xi ^2}{x^2}{y^2} + 78{x^2}{y^2} - 48{x^2} - \\
 - {y^6} + {\xi ^2}{y^4} + 18\xi {y^4} - 15{y^4} - 16{\xi ^3}{y^2} - 80{\xi ^2}{y^2} + 144\xi {y^2} - 48{y^2} + 64=0.
\end{multline}
Thus in the regions ${Q_0}(\xi)$, ${Q_2}(\xi)$, ${Q_3}(\xi )$ the discriminant $D(x,y,\xi )>0$, while $D(x,y,\xi )<0$ in ${Q_1}(\xi)$, and $D(x,y,\xi )=0$ on the critical curve $\Gamma(\xi)$.  Figure~\ref{fig45} depicts $\Gamma (\xi )$ on the plane of slow variables for different values of $\xi$. At $\left| {\,\xi \,} \right| < 3$ critical curve have cusps, which lie on astroid \eqref{astroid}. Additionally this curve may have points of self-intersection. If $-3<\xi<-1$, the curve intersects itself on axis $x$, with the $x$ coordinates of self-intersection points being defined by equation
\[{x^2} + 8(\xi  + 1) = 0.\]
If $1<\xi<3$, points of self-intersection lie on $y$ axis, and their $y$ coordinates are defined by equation
\[{y^2} - 8(\xi  - 1) = 0.\]

The critical curve allows a parametrization
\begin{equation}\label{(12)}
\Gamma (\xi ) = \left\{ \left. x = \cos \phiarch (\xi  + \cos 2 \phiarch  - 2),~
y = \sin \phiarch (\xi  + \cos 2 \phiarch  + 2)~\right|~\phiarch  \in {S^1} \right\},
\end{equation}
which is illustrated by Figure~\ref{fig45}. The parameter $\phiarch$ in \eqref{(12)} coincide with the critical points of potential $W(\varphi)$ at given level $\xi$, as depicted in Figure~\ref{fig3}, in respective points of $(x,y)$ plane:
\[
\begin{cases}
W(\phiarch ,x(\phiarch ,\xi ),y(\phiarch ,\xi )) = \xi,\\
\left.\frac{{\partial}}{{\partial \varphi }}W(\varphi,x(\phiarch,\xi ),y(\phiarch ,\xi ))\right|_{\varphi=\phiarch} = 0.
\end{cases}
\]

The introduced parametric representation is convenient, in particular, for defining the location of cusps and self-intersection points. For cusps $\phiarch  = {\varphi ^*}$, where ${\varphi ^*}$ is obtained from the equation $\tan^2{\varphi ^*} = {{(3 + \xi )} \mathord{\left/ {\vphantom {{(3 + \xi )} {(3 - \xi )}}} \right. \kern-\nulldelimiterspace} {(3 - \xi )}}$, which have four roots on $S_1$. We shall denote these cusps as $Y_1$,...,$Y_4$ with the lower index being the number of a quadrant, in which the respective value $\varphi ^*$ lies. Self-intersection points of $\Gamma (\xi )$ on $x$ axis ($-3<\xi<-1$) we shall denote as $B_1$ and $B_2$ for right and left half-planes respectively. Self-intersection points on $y$ axis ($1<\xi<3$) we shall denote as $S_1$ and $S_2$ for upper and lower half-planes.

\emph{Note:} It can be demonstrated, that curves $\Gamma(\xi )$ are the involutes of astroid \eqref{astroid} constructed with tethers of length $3\pm\xi$ extended from astroid's cusps.
This makes $\Gamma (\xi )$ also a family of equidistant curves with the distance $\left| {{\xi _a} - {\xi _b}} \right|$ between any two curves $\Gamma (\xi_a)$ and $\Gamma (\xi_b)$.

\subsection{Transformation of regions ${Q_0}(\xi ),...,{Q_3}(\xi )$ with change of $\xi$}
It should be noted first, that region with a single librating solution ${Q_1}(\xi )$ is present on plane $(x,y)$ for all values of $\xi$. Other regions appear and disappear, as $\xi$ crosses several bifurcation values $\xi_i$:
\[
\xi_1=-3,~\xi_2=-1,~\xi_3=1,~\xi_4=3.
\]
We shall describe, how the regions are transformed, as $\xi$ increases.

If $\xi<\xi_1$, there exists a forbidden region ${Q_0}(\xi )$ around the point $(0,0)$, with the rest of $(x,y)$ plane being the ${Q_1}(\xi )$ region.

\begin{figure}[ht]
\center{\includegraphics[width=122mm]{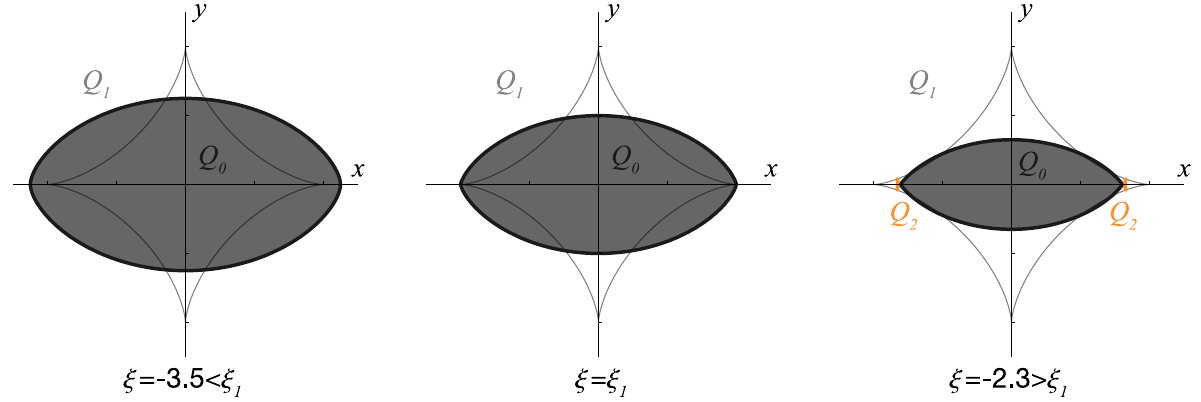}}
\caption{Bifurcation at $\xi=\xi_1$: appearance of region $Q_2$. Here and further region $Q_0$ is colored dark gray, region $Q_2$ -- orange. The rest blank space corresponds to region $Q_1$}
\label{fig6}
\end{figure}

At $\xi=\xi _1$ on the right and on the left from region ${Q_0}(\xi )$ two parts of region ${Q_2}(\xi )$ (region with two librating solutions) appear~(Fig.~\ref{fig6}).

\begin{figure}[ht]
\center{\includegraphics[width=122mm]{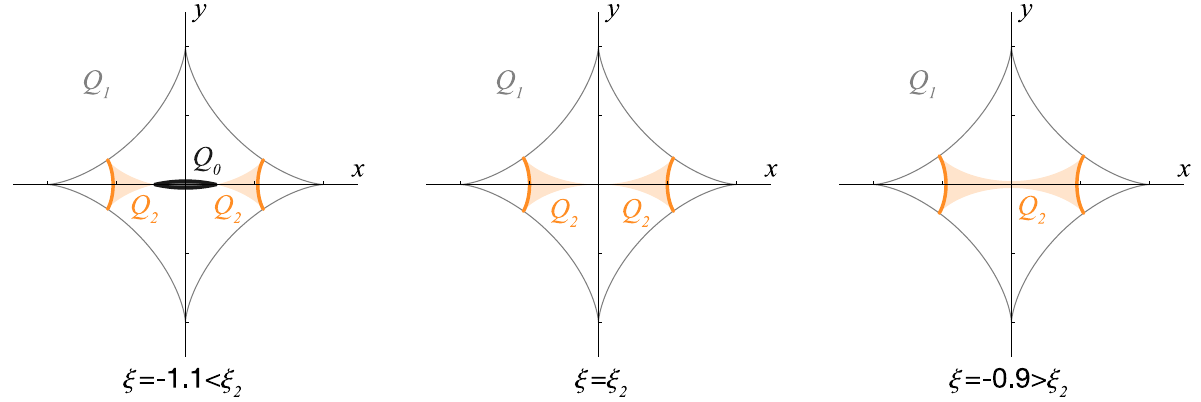}}
\caption{Bifurcation at $\xi=\xi_2$: vanishing of forbidden region $Q_0$}
\label{fig7}
\end{figure}

At $\xi=\xi_2$ the region ${Q_0}(\xi )$ disappears and ${Q_2}(\xi )$ becomes connected~(Fig.~\ref{fig7}).

\begin{figure}[H]
\center{\includegraphics[width=122mm]{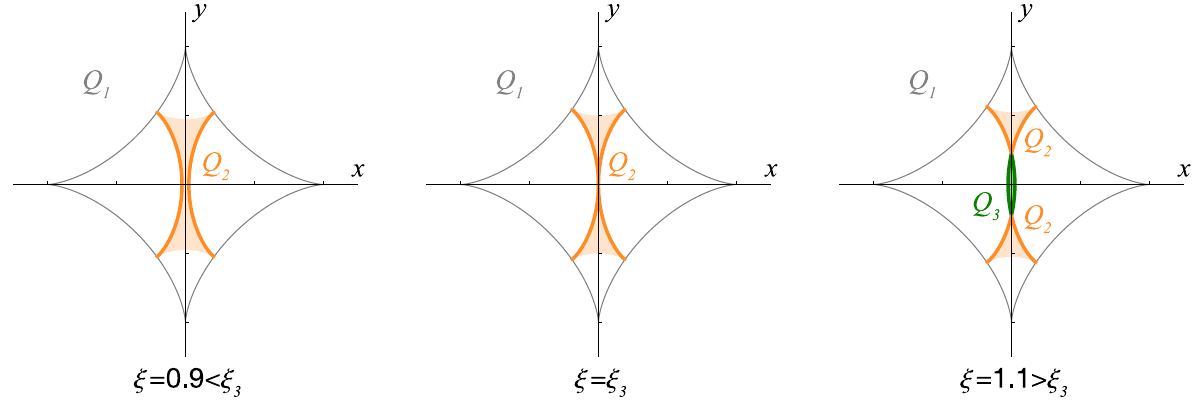}}
\caption{Bifurcation at $\xi=\xi_3$: appearance of region $Q_3$ (here and further colored green)}
\label{fig8}
\end{figure}

At $\xi=\xi_3$ region ${Q_2}(\xi )$ is separated into two parts again by appearing region ${Q_3}(\xi )$ (region with circulating resonant angle) around the point $(0,0)$ as seen in Figure~\ref{fig8}.

\begin{figure}[ht]
\center{\includegraphics[width=122mm]{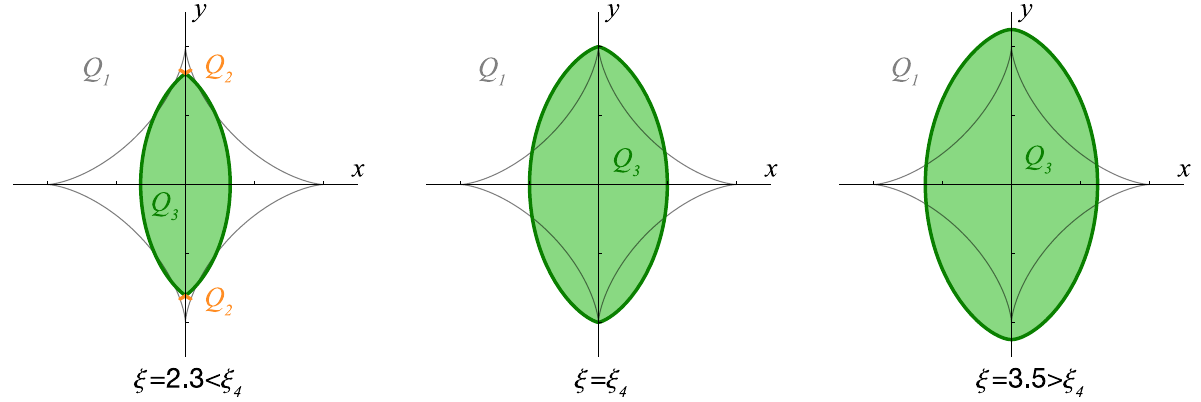}}
\caption{Bifurcation at $\xi=\xi_4$: vanishing of region $Q_2$}
\label{fig9}
\end{figure}

At $\xi=\xi_4$ region ${Q_2}(\xi )$ disappears (Fig.~\ref{fig9}). For $\xi>\xi_4$ there exists only region ${Q_3}(\xi )$ surrounded by ${Q_1}(\xi )$.
	
Let us now describe how borders of regions ${Q_0}(\xi ),...,{Q_3}(\xi )$ transforms with increase of $\xi$. The border between ${Q_0}(\xi )$ and ${Q_1}(\xi )$ we shall call the \emph{existence curve} and denote it as ${\Gamma _0}(\xi )$. It corresponds to the part of the critical curve $\Gamma(\varepsilon)$ in which ${W_{min}}(x,y)=\xi$. For $\xi<{\xi _1}$ curve $\Gamma(\varepsilon)\equiv\Gamma_0(\varepsilon)$. For ${\xi _1}<\xi< {\xi _2}$ curve $\Gamma_0(\varepsilon)$ consists of two intervals of $\Gamma(\varepsilon)$, lying between points of self-intersection $B_1$ and $B_2$.

We shall adopt the traditional terminology common in studies of slow-fast systems \citep{Wisdom85, Neishtadt_kirkwood} with modifications made to better represent the specifics of the discussed problem. The border between regions ${Q_1}(\xi)$ and ${Q_3}(\xi)$ we shall call an \emph{uncertainty curve of the first kind} and use a notation ${\Gamma _1}(\xi)$ for it. Points of the uncertainty curve of the first kind are defined by condition ${W_{\max}}(x,y)=\xi$. If $\xi  > {\xi _4}$, then ${\Gamma_1}(\xi)\equiv\Gamma (\xi )$. For ${\xi_3}<\xi<{\xi_4}$, the curve ${\Gamma_1}(\xi)$ consists of ${\Gamma}(\xi)$ parts, which are contained between points of self-intersection ${S_1}$ and ${S_2}$.

\begin{figure}[ht]
\center{\includegraphics[width=80mm]{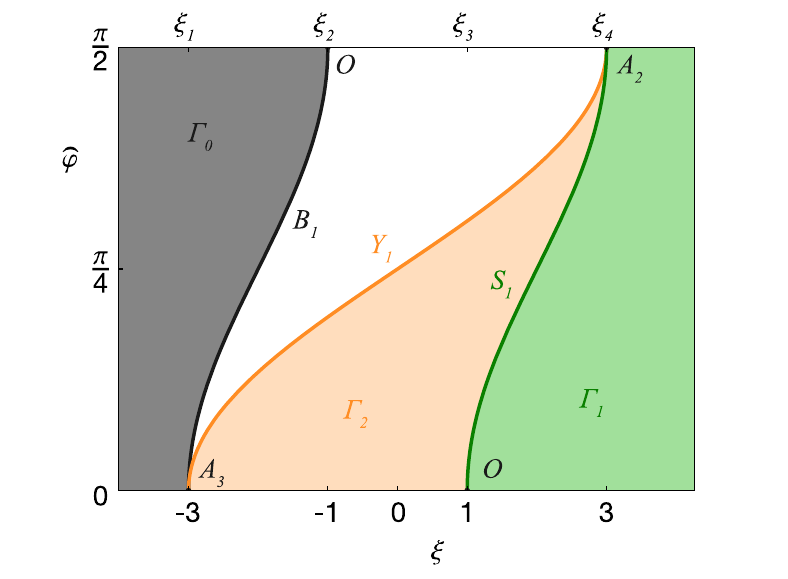}}
\caption{Diagram showing the partition of critical curve into existence curve $\Gamma_0(\xi)$ and critical curves $\Gamma_{1,2}(\xi)$}
\label{fig10}
\end{figure}

The part of the border between ${Q_1}(\xi )$ and ${Q_3}(\xi)$, along which holds the equality $W_{\max}^*(x,y)=\xi$, we shall call an \emph{uncertainty curve of the second kind} and denote it ${\Gamma_2}(\xi)$. For ${\xi_1}<\xi<{\xi_3}$ the ${\Gamma _2}(\xi ) = {Y_1}{Y_3} \cup {Y_2}{Y_4}$, where ${Y_1}{Y_3}$ and ${Y_2}{Y_4}$ are segments of $\Gamma(\xi)$, which lie between corresponding cusps. If ${\xi _1} < \xi  < {\xi _2}$ then curve ${\Gamma _2}(\xi ) = {S_1}{Y_1} \cup {S_1}{Y_2} \cup {S_2}{Y_3} \cup {S_2}{Y_4}$.
For the rest part of the border between ${Q_1}(\xi )$ and ${Q_3}(\xi)$ holds $W_{\min }^*(x,y) = \xi $. As no dynamical effects of interest are happening on this segment, we shall not refer to it further.

Figure~\ref{fig10} depicts a diagram, that shows values of $\phiarch$ defining positions of cusps and self-intersection points on $\Gamma (\xi )$, as well as the segments which correspond to existence curve and two uncertainty curves. Due to the symmetry, it is sufficient to consider only $\phiarch \in \left[0,\pi/2\right]$ (Fig.~\ref{fig45}).

\subsection{Three-dimensional representation of the set of curves $\Gamma (\xi )$}

\begin{figure}[H]
\center{\includegraphics[width=122 mm]{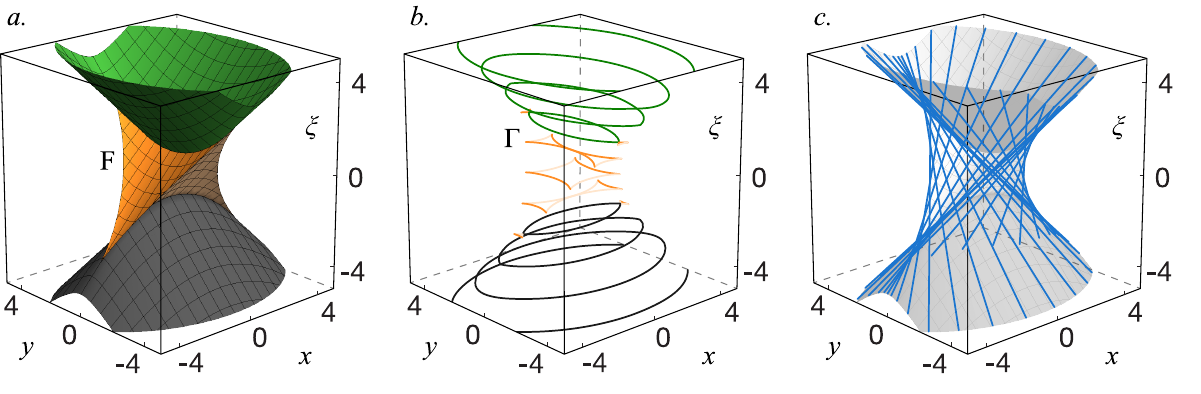}}
\caption{Surface $F$ composed of curves $\Gamma(\xi)$ in $xy\xi$ space: a. general representation of the surface, b. surface's cross sections by equi-Hamiltonian planes, c. rulings of surface $F$}
\label{fig11}
\end{figure}

Curves $\Gamma (\xi )$ can be interpreted as cross sections of some surface $F$ in the space $xy\xi$ by equi-Hamiltonian planes $\xi  = const$ (Fig.~ \ref{fig11}a,b). In this space for fixed value of $\phiarch $ the equations \eqref{(12)} define a straight line, which means that the surface $F$ is ruled~(Fig.~\ref{fig11}c).

The same surface $F$ defined by the equation analogous to \eqref{(11)} also appears in a completely different problem studied by \cite{Batkhin}, where it partitions the parametric space of some mechanical system into regions with different stability properties.

\section{Evolution equations construction}\label{Sec:slow}

\subsection{Averaging along fast subsystem's solutions}

Considering that
\[
\frac{{\partial \Xi }}{{\partial x}} = \cos \varphi ,\quad \frac{{\partial \Xi }}{{\partial y}} = \sin \varphi,
\]
construction of the evolution equations \eqref{(4)} require calculating two averaged properties:
\begin{equation}\label{(13)}
\begin{array}{c}
{\left\langle {\sin \varphi } \right\rangle  = \frac{1}{{T(x,y,\xi )}}\int\limits_0^{T(x,y,\xi )} {\sin \varphi (\tau,x,y,\xi ){\mkern 1mu} {\kern 1pt} d} \tau,}\\
{\left\langle {\cos \varphi } \right\rangle  = \frac{1}{{T(x,y,\xi )}}\int\limits_0^{T(x,y,\xi )} {\cos \varphi (\tau,x,y,\xi ){\mkern 1mu} {\kern 1pt} d} \tau.}
\end{array}
\end{equation}
The equality
\[
\frac{{d\varphi }}{{d\tau}} = \Phi  =  \pm \sqrt {2\left[ {\xi  - W(x,y,\varphi )} \right]}
\]
allows finding a period of fast subsystem's solution at Hamiltonian level $\Xi =\xi$ and calculating (after proper change of variables) values of integrals on the right-hand side of \eqref{(13)}, e.g. for librating solutions:
	
\begin{equation}\label{(14)}
\begin{array}{c}
{T(x,y,\xi ) = 2\int\limits_{{\varphi _*}}^{{\varphi ^*}} {\frac{{d\varphi }}{{\sqrt {2\left[ {\xi  - W(x,y,\varphi )} \right]} }}} ,}\\
{\int\limits_0^{T(x,y,\xi )} {f\left(\varphi (t,x,y,\xi )\right) d\tau = 2\int\limits_{{\varphi _*}}^{{\varphi ^*}} {\frac{f(\varphi )d\varphi}{{\sqrt {2\left[ {\xi  - W(x,y,\varphi )} \right]} }}} } \,.}
\end{array}
\end{equation}

Here $\varphi _*$ and $\varphi^*$ denote minimum and maximum values of angle $\varphi$ in librating solution, along which the averaging is being performed.

For the system \eqref{(2)} period $T(x,y,\xi )$ and integrals \eqref{(13)} can be expressed in terms of complete elliptical integrals of the first and the third kind. Further the concise description of this transformation is given, using the case
\begin{equation}\label{(15)}
- \pi  \le {\varphi _*} < {\varphi ^*} \le \pi
\end{equation}
as an example. After standard substitution $\lambda=\tan\left(\varphi/2\right)$ we obtain

\begin{equation}\label{(16)}
\begin{array}{c}
T(x,y,\xi ) = 4\int\limits_{\lambda _*}^{\lambda^*}
 \frac{d\lambda}{\sqrt {2{R_4}(\lambda)}},
 \\
\left\langle {\sin \varphi } \right\rangle={\int\limits_{\varphi _*}^{\varphi ^*} {\frac{\sin \varphi d\varphi }{\sqrt {2\left[ {\xi  - W(x,y,\varphi)} \right]} } = } 4\int\limits_{\lambda _*}^{\lambda^*} {\frac{\lambda d\lambda}{{(1 + {\lambda ^2})\sqrt {2{R_4}(\lambda )} }}} ,}\\
\left\langle {\cos \varphi } \right\rangle={\int\limits_{{\varphi _*}}^{{\varphi ^*}} {\frac{{\cos \varphi d\varphi }}{{\sqrt {2\left[ {\xi  - W(x,y,\varphi )} \right]} }} = } 2\int\limits_{\lambda _*}^{\lambda^*} {\frac{{\left(1 - \lambda^2\right)d\lambda }}{{(1 + {\lambda ^2})\sqrt {2{R_4}(\lambda )} }}} .}
\end{array}
\end{equation}
Here $\lambda _* = \tan\left(\varphi_*/2\right)$, $\lambda^*=\tan\left(\varphi^*/2\right)$. Function $R_4\left(\lambda\right)$ in \eqref{(16)} is a quartic polynomial
\[
{R_4}(\lambda ) = {d_0}{\lambda ^4} + {d_1}{\lambda ^3} + {d_2}{\lambda ^2} + {d_3}\lambda  + {d_4}
\]
with coefficients
\[
{d_0} = \xi  - 1 + x,\quad
{d_1} = {d_3} =  - 2y,\quad
{d_2} = 2\xi  + 6,\quad
{d_4} = \xi  - 1 - x.
\]

Integrals on the right-hand side in \eqref{(16)} can be rewritten as follows:
\begin{equation}\label{(17)}
\begin{array}{l}
T(x,y,\xi ) = \frac{4}{{\sqrt {2\left| {{d_0}} \right|} }}{I_{0,0}},
\\
\int\limits_{{\varphi _*}}^{{\varphi ^*}} {\frac{{\sin \varphi d\varphi }}{{\sqrt {2\left[ {\xi  - W(x,y,\varphi )} \right]} }} = } \frac{4}{{\sqrt {2\left| {{d_0}} \right|} }}{I_{1,1}},
\\
\int\limits_{{\varphi _*}}^{{\varphi ^*}} {\frac{{\cos \varphi d\varphi }}{{\sqrt {2\left[ {\xi  - W(x,y,\varphi )} \right]} }} = } \frac{2}{{\sqrt {2\left| {{d_0}} \right|} }}\left( {2{I_{1,0}} - {I_{0,0}}} \right),
\end{array}
\end{equation}
where notation ${I_{k,r}}$ is used for integrals:
\begin{equation}\label{(18)}
I_{k,r} = \int\limits_{{\lambda _*}}^{{\lambda ^*}} {\frac{{{\lambda ^r}d\lambda }}{{{{({\lambda ^2} + 1)}^k}\sqrt { \pm (\lambda  - {a_1})(\lambda  - {a_2})(\lambda  - {a_3})(\lambda  - {a_4})} }}}.
\end{equation}
$a_k$ -- roots of polinomial $R_4(\lambda)$, and the sign in the square root is determined by sign of coefficient $d_0$. Integrals \eqref{(18)} can be presented as linear combinations of elliptic integrals of the first and the third kind:
\begin{equation}\label{(19)}
\begin{array}{l}
I_{0,0} = {c_{0,0}}K(k),
\\
I_{1,0} = {c_{1,1}}K(k) + {c_{1,3}}\Pi \left( {h,k} \right) + {{\bar c}_{1,3}}\Pi \left( {\bar h,k} \right),
\\
I_{1,1} = {g_{1,1}}K(k) + {g_{1,3}}\Pi \left( {h,k} \right) + {{\bar g}_{1,3}}\Pi \left( {\bar h,k} \right).
\end{array}
\end{equation}
Analytical expressions for coefficients $c_{m,l}$, $g_{m,l}$, moduli $k$, and parameters $h$ in \eqref{(19)} depend on integration interval and properties of $R_4(\lambda)$ roots. These expressions are gathered in the Appendix B.

Expressions of the kind
\[
c\Pi (h,k) + \bar c\Pi (\bar h,k)\quad (c,h \in \mathbb{C},~ k \in \mathbb{R},~ 0 < {k^2} < 1)
\]
can be further reduced to linear combinations of complete elliptic integrals of the first kind and the third kind with real parameter \citep{ByrdFriedman, Lang60}. This, however, results in more complicated expressions, which is why we use representation \eqref{(19)}.

When averaging along librating solutions of fast subsystem, which violate the condition \eqref{(15)}, after substitution $\lambda=\tan\left(\varphi/2\right)$ the integration in \eqref{(16)} is carried over two semi-infinite intervals. E.g., for $\varphi _*<\pi,~ \varphi^*>\pi,~ \varphi ^*-\varphi _*<2\pi$ the expression for the period $T$ is
\[
T(x,y,\xi ) = 4\left( {\int\limits_{-\infty}^{\lambda^*} {\frac{{d\lambda }}{{\sqrt {2{R_4}(\lambda )} }}}  + \int\limits_{\lambda_*}^{+\infty} {\frac{{d\lambda }}{{\sqrt {2{R_4}(\lambda )} }}} } \right).
\]
In these cases it is implied that all $I_{k,r}$ in \eqref{(17)} are the sum of two integrals as well.

After all described transformations evolution equations \eqref{(4)} take a simple form
\begin{equation}\label{(20)}
\frac{dx}{d\tau} = \varepsilon \frac{2 I_{1,1}}{I_{0,0}},\quad
\frac{dy}{d\tau} = \varepsilon \left[1 - \frac{2 I_{1,0}}{I_{0,0}}\right].
\end{equation}

It should be noted, that there is an ambiguity in calculation of the right-hand side parts of evolution equations in the region ${Q_2}(\xi)$: the result depends on the choice of the fast subsystem's solution, and there are two different librating solutions in ${Q_2}(\xi)$. Consequently, phase portraits of \eqref{(4)} shall contain two sets of trajectories in ${Q_2}(\xi)$, which correspond to two possible variants of slow variables' evolution.

When describing the crossing of uncertainty curves $\Gamma_i(\xi)$ by the projection $\boldsymbol{\zeta}(\tau)=(x(\tau),y(\tau))^T$ of the phase point ${\bf{z}}(\tau) = {(\varphi (\tau),\Phi (\tau),x(\tau),y(\tau))^T}$ on the plane $(x,y)$, we shall confine ourselves to formal continuation of averaged system's trajectories, lying on opposite sides of $\Gamma_i$. The detailed analysis of these events is given in \cite{Neishtadt_kirkwood}, \cite{Neishtadt04}, and \cite{Sidorenko_QS}.

\subsection{Fast subsystem's action variable -- integral of evolution equations}
As was mentioned in the Section~\ref{Sec:modelH}, adiabatic invariant $J(x,y,\xi)$ of slow-fast system is a first integral of evolution equations~\eqref{(4)}. Formula \eqref{(6)} for $J(x,y,\xi)$ can be expressed as a linear combination of integrals~$I_{k,r}$ defined by \eqref{(18)}:
\begin{multline}\label{(21)}
J(x,y,\xi ) = \frac{1}{\pi }\int\limits_{{\varphi _*}}^{{\varphi ^*}} {\Phi d\varphi  = } \frac{1}{\pi }\int\limits_{{\varphi _*}}^{{\varphi ^*}} {\sqrt {2\left[ {\xi  - W(x,y,\varphi )} \right]} d\varphi  = } \frac{2}{\pi }\int\limits_{{\lambda _*}}^{{\lambda ^*}} {\frac{{\sqrt {2{R_4}(\lambda )} d\lambda }}{{{{\left( {1 + {\lambda ^2}} \right)}^2}}} = } \\
 = \frac{{2\sqrt 2 }}{{\pi \sqrt {\left| {{d_0}} \right|} }}\left[ {{d_0}{I_{0,0}} + {d_1}{I_{1,1}} + ({d_2} - 2{d_0}){I_{1,0}} + ({d_4} + {d_0} - {d_2}){I_{2,0}}} \right].
\end{multline}
Analytical expression for $I_{2,0}$ is presented in the Appendix B alongside the rest of the integrals previously shown in~\eqref{(19)}.

\begin{figure}[H]
\includegraphics[width=122mm]{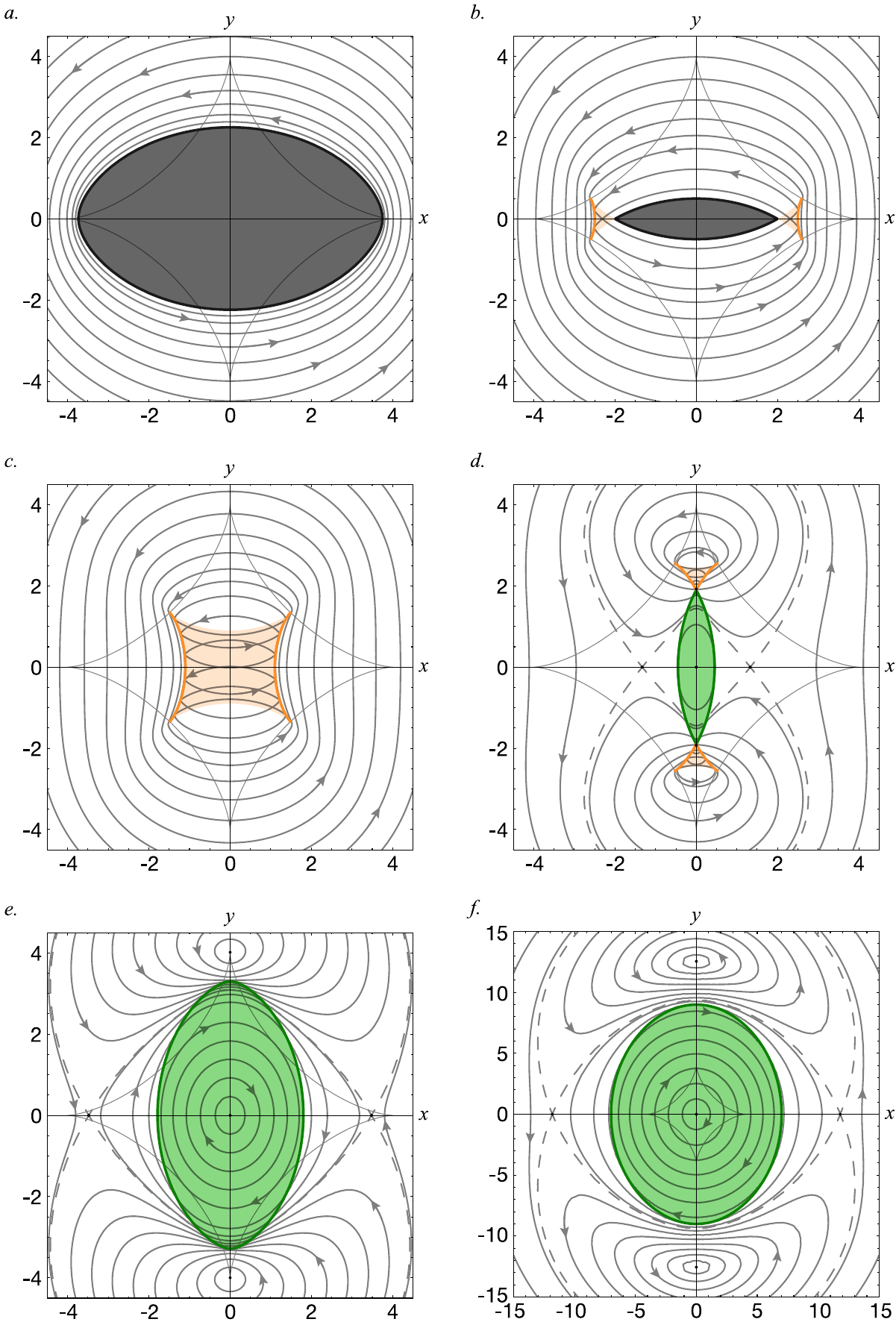}
\caption{Phase portraits of evolution equations: \newline
a.~$\xi=-4$, b.~$\xi=-1.5$, c.~$\xi=-0.1$, d.~$\xi=1.45$, e.~$\xi=3.4$, f.~$\xi=8.0$}
\label{fig12}
\end{figure}

\section{Study of slow variables' behavior using evolution equations}\label{Sec:PP}

\subsection{Phase portraits of evolution equations. Stationary points}

To analyze solutions of the slow subsystem \eqref{(4)}, we build its phase portraits. For values $\xi<\xi_1$ the structure of phase portrait is simple -- all phase trajectories are represented by closed loops encircling the forbidden region ${Q_0}(\xi)$ (Fig.~\ref{fig12}a). Figure~\ref{fig12}b depicts a typical phase portrait for $\xi \in (\xi_1,\xi_2)$ -- two symmetrical parts of region ${Q_2}(\xi)$ adjoining the central region ${Q_0}(\xi)$ contain two layers of phase trajectories. For $\xi \in(\xi_2,\xi_3)$ there are only two regions on the phase plane, i.e. $Q_0(\xi )$ and $Q_2(\xi)$ (Fig.~\ref{fig12}c). In addition to presented in Figure~\ref{fig12}c,d change of phase portrait's global structure, the behaviour of phase trajectories near the uncertainty curves $\Gamma_{1,2}(\xi)$ at $\xi \in(\xi_2,\xi_4)$ have some specific qualitative differences at different $\xi$ values. The detailed description of that is given in the end of this section.

Phase portraits at $\xi>\xi_3$ have five stationary points: the origin of $xy$-plane is the stable point of the center type, two more center points are symmetrically located on the $y$-axis, and two unstable saddle points are symmetrically located on the $x$-axis (Fig.~\ref{fig12}d--f). Phase portraits depicted in Figures~\ref{fig12}e and \ref{fig12}f are differ in relative positions of heteroclinic trajectories and uncertainty curve $\Gamma_1$.

Ordinates $y$ of the center points above and below plane's origin are defined by equation
\begin{equation}\label{(22)}
K\left( m \right) = 2\Pi \left( {n\left| m\right.} \right),
\end{equation}
where
\[
m = \frac{U_+}{U_-},\quad
n = \frac{U_+}{\xi+1-|y|},\quad
U_\pm= \xi-3\pm\sqrt{y^2+8(1-\xi)}.
\]
Abscissae $x$ of the saddle points are defined by the same relation \eqref{(22)} with different parameters:
\[
m=\frac{Q_{+ -}}{Q_{- -}}\frac{Q_{+ +}}{Q_{- +}},\quad
n=\frac{Q_{+ -}}{Q_{- -}},\quad
Q_{\pm \pm}=\sqrt{x^2+8(1+\xi)}\pm 4\pm |x|.
\]

\begin{figure}[H]
\center{
\includegraphics[width=122mm]{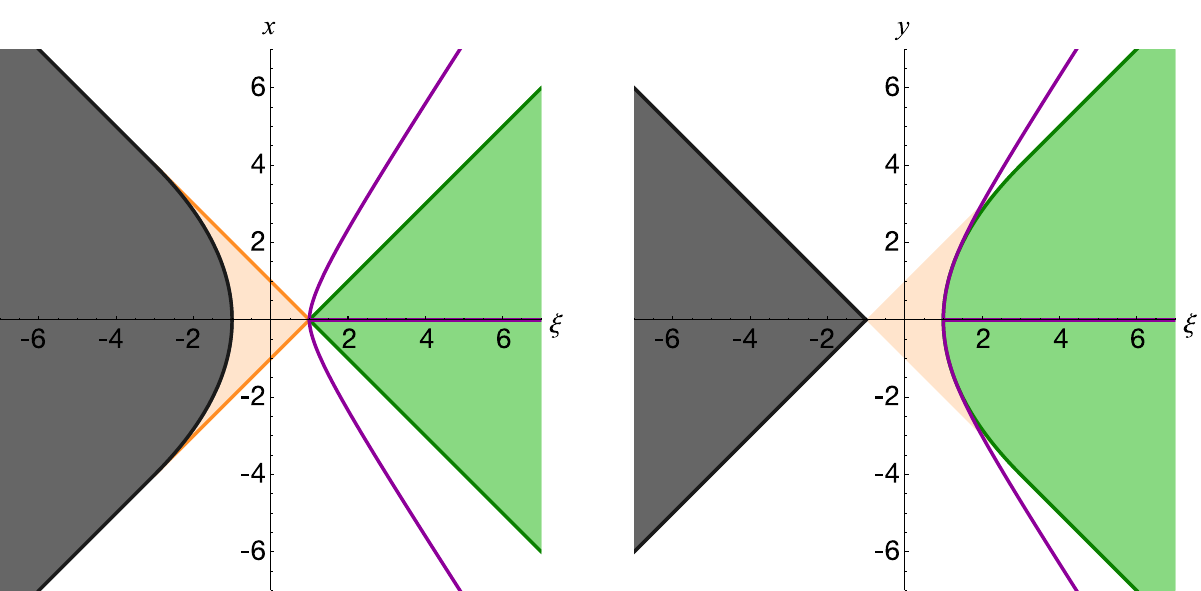}
}
\caption{Coordinates of evolution equations' stationary points, which lie on axis $x$ (left), and on axis $y$ (right). Values of coordinates as functions of $\xi$ are represented by violet lines. The rest of the coloring is consistent with Figure~\ref{fig12} in denoting the regions $Q_i$ and points from different parts of the critical curve}
\label{fig13}
\end{figure}

\begin{figure}[H]
\center{\includegraphics[width=61mm]{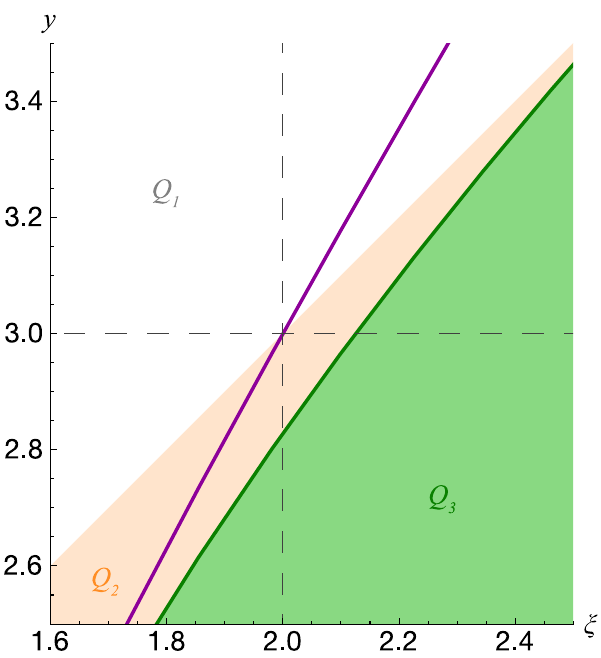}}
\caption{Transition of the top center point from $Q_2(\xi)$ to $Q_1(\xi)$ at $\xi=\xi_5=2$}
\label{fig14}
\end{figure}

Solutions of~\eqref{(22)} are plotted in Figure~\ref{fig13} as functions of $\xi$ for both types of stationary points. Note, that top and bottom center points are located in $Q_2$ at $\xi<\xi_5=2$ and in $Q_1$ at $\xi>\xi_5$ (Fig.~\ref{fig14}).

\subsection{Limiting points on uncertainty curves}

To conclude the description of how phase portraits' topology changes with $\xi$ value, points of uncertainty curves' intersections with limiting trajectories, that are limiting cases for different families of trajectories, should be considered. We shall refer to them as \emph{limiting points}.
\begin{figure}[H]
\center{\includegraphics[width=80mm]{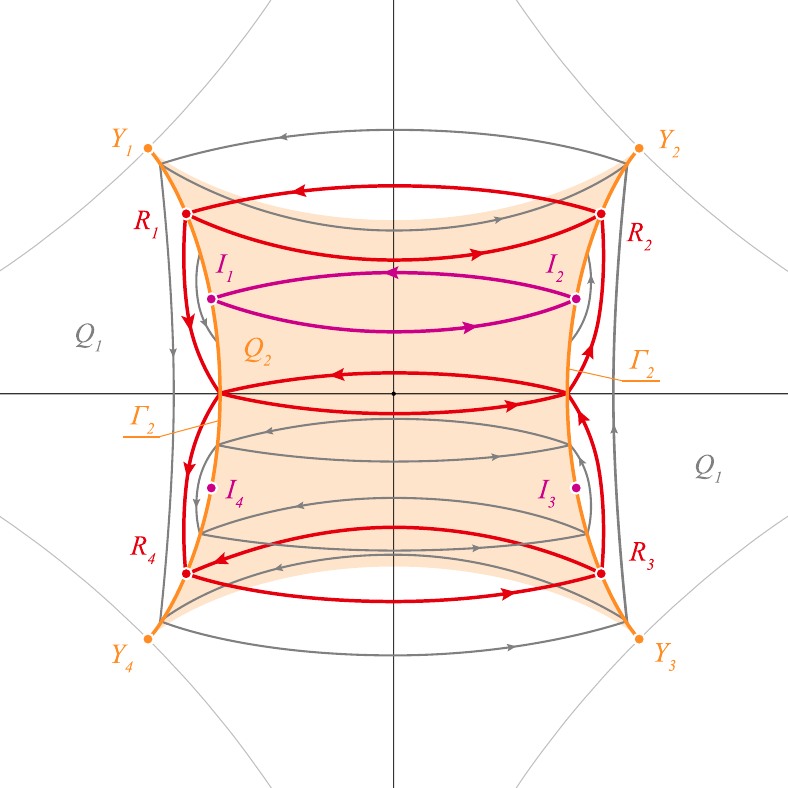}}
\caption{Limiting points $R_i$ and $I_i$ at $\xi\approx0$}
\label{fig15}
\end{figure}
Several types of limiting points are depicted in Figure~\ref{fig15}. Trajectories, which go in $Q_1$ between two points on uncertainty curve $\Gamma_2$, can be divided into two families: the trajectories that intersect $x$ axis, and those that do not. Thus there is limiting trajectory that separates these two families (in Figure~\ref{fig15} it is colored red). We shall denote the limiting points corresponding to this trajectory as $R_1,...,R_4$ (Fig.~\ref{fig15}).
The subset of trajectories, which do not cross $x$-axis in $Q_1$ as another limiting case contain trajectories, which come to uncertainty curve from the side of $Q_2$ and then reflect back without exiting to $Q_1$. In Figure~\ref{fig15} these trajectories are colored purple, and the corresponding limiting points are denoted $I_1,...,I_4$.

\begin{figure}[H]
\center{\includegraphics[width=122mm]{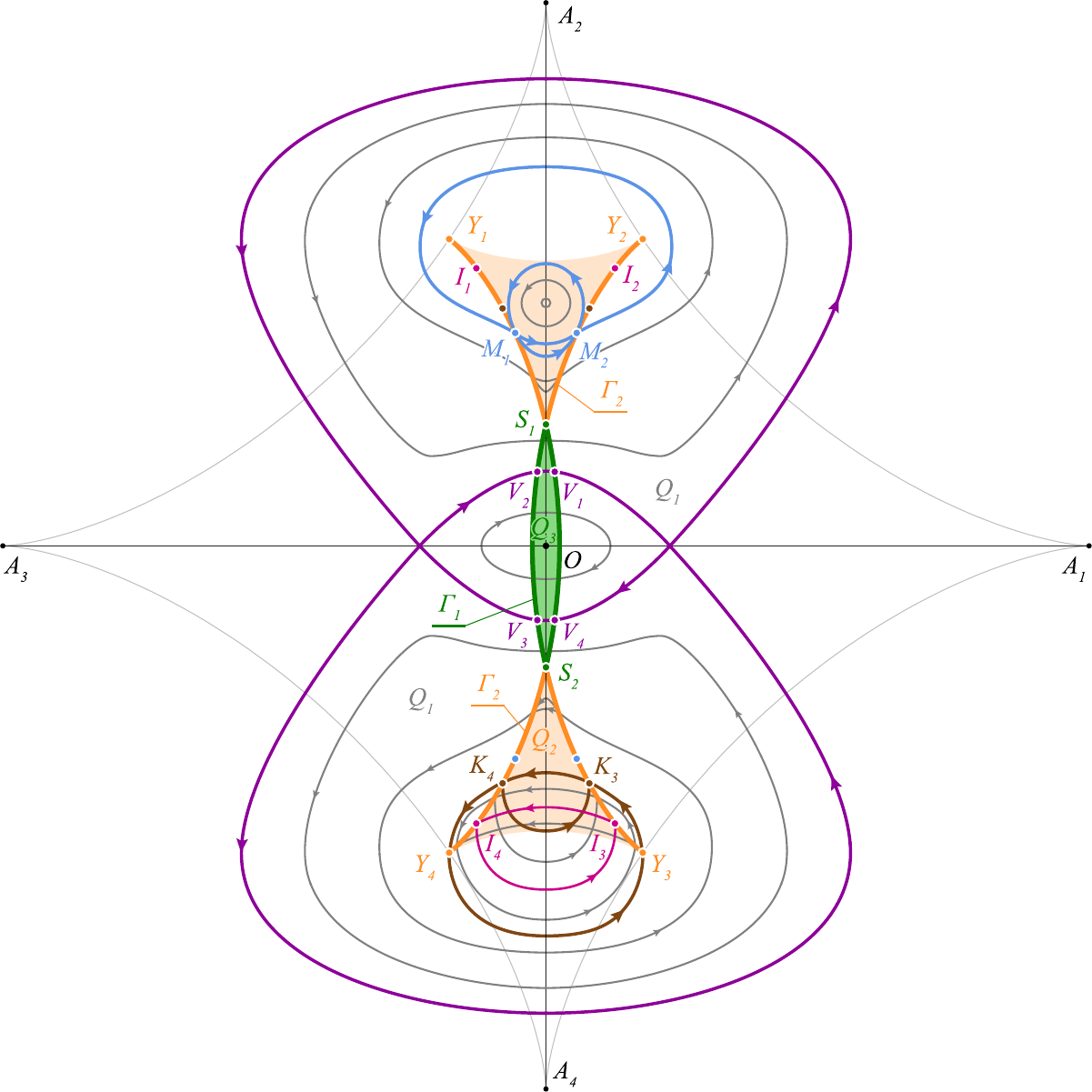}}
\caption{Limiting points $K_i$, $M_i$ and $V_i$ ($\xi\sim 1.5$)}
\label{fig16}
\end{figure}

More limiting points are depicted in figure~\ref{fig16}. Points ${K_1},...,{K_4} \in {\Gamma _2}$ are connected to cusps $Y_1,...,Y_4$ by limiting phase trajectories, which separate the family of trajectories lying to the one side of uncertainty curve $\Gamma_2$ from those which in $Q_1$ connect symmetric points on left and right sides of uncertainty curves $\Gamma_{1,2}$.
Limiting points ${M_1},...,{M_4}$ divides each of four $\Gamma_2$ segments (located in each of four quadrants of $xy$ plane) into two sections: one section has both trajectories, which adjoin from $Q_2$ side, going in the same direction, while trajectories adjoining the other section go in opposite directions (also see Figure~\ref{fig19}).
Points ${V_1},...,{V_4}$ of $\Gamma_1$ curve's intersections with mentioned earlier separatrices constitute the last type of limiting points.

\begin{figure}[H]
\center{\includegraphics[width=80mm]{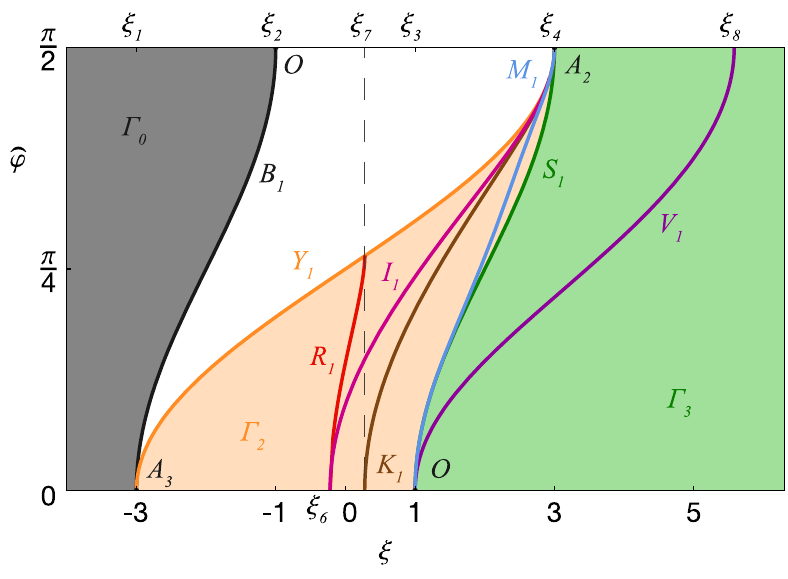}}
\caption{Bifurcation diagram showing the dependance of limiting points position on the critical curve $\Gamma$ on $\xi$.}
\label{fig17}
\end{figure}

All differences between the phase portraits (Fig.~\ref{fig12}) emerge from changes in position of limiting points on the critical curve. By adding these points to diagram in Figure~\ref{fig10} we obtain a bifurcation diagram (Fig.~\ref{fig17}), from which all changes in topological structure of phase portraits can be understood.
Let us describe, what is happening to different limiting points by going successively from low to high values of $\xi$.
The first limiting points -- $R_i$ and $I_i$ -- appear at $\xi=\xi _6\approx -0.22073$. At $\xi=\xi_7\approx 0.27704$ points $R_i$ merge with cusps $Y_i$, and at the same time points $K_i$ appear. Points $M_i$ and $V_i$ emerge simultaneously with region $Q_3$ and self-intersections points $S_i$ of the critical curve at $\xi=\xi_3$. All points on the $\Gamma_2$ part of the critical curve -- $I_i$, $K_i$, and $M_i$, as well as the ends $Y_i$, $S_i$ of $\Gamma_2$ itself -- disappear at $\xi  = {\xi _4}$ merging with the astroid's cusps $A_2$ and $A_4$. Finally, at $\xi=\xi_8\approx 5.57954$ points $V_i$ disappear by merging pairwise.

Bifurcation Hamiltonian values $\xi_1,\ldots,\xi_8$ partition the whole range of $\xi$ into nine intervals, meaning that there is a total number of nine different types of phase portraits for slow subsystem. The bifurcation at $\xi_5$ (Fig.~\ref{fig14}) is omitted from Figure~\ref{fig17} because it does not bear any significance for the dynamical effects considered further.

\section{Quasi-random effects}\label{Sec:chaos}

\subsection{Probabilistic change of fast subsystem's motion regime on the uncertainty curve of the second kind}
\begin{figure}[H]
\center{
\includegraphics[width=80mm]{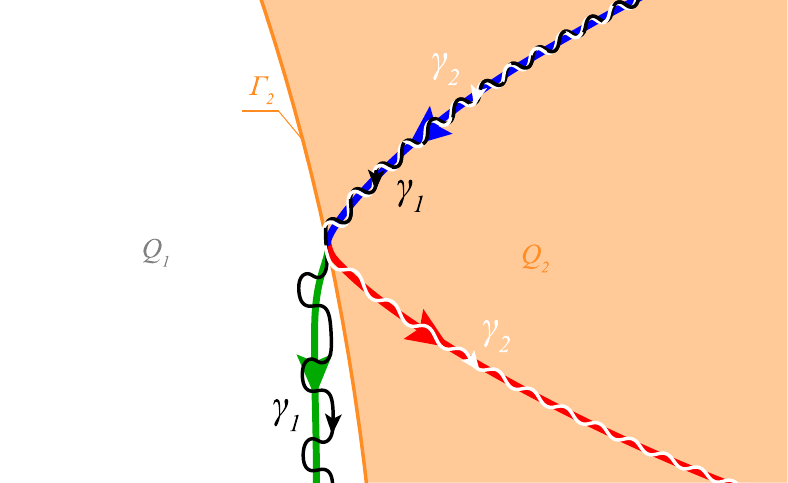}
}
\caption{Two phase trajectories starting from very close initial conditions may diverge at uncertainty curve $\Gamma_2$, as $\gamma_1$ and $\gamma_2$ do. Green and red lines show two guiding trajectories departing from the same point of $\Gamma_2$, to which the blue one arrives. Trajectory $\gamma$ of non-averaged system may go along either one of them after reaching the border between $Q_1$~and~$Q_2$}
\label{figpre18}
\end{figure}

In each point of ${\Gamma _2}(\xi )$ three \emph{guiding trajectories}, i.e. trajectories of averaged system \eqref{(4)}, meet: two on the side of ${Q_2}(\xi )$ region and one on the ${Q_1}(\xi )$ side. In the case, when two out of these three trajectories are outgoing, the transition of the phase point to either one of them can be considered as a probabilistic event. In the original system \eqref{(1)} initial values of fast variables corresponding to two different outcomes are strongly mixed in the phase space. Therefore even small variation of initial conditions ${\bf{z}}(0) = {(\varphi (0),\Phi (0),x(0),y(0))^T}$ can lead to qualitative change in system's evolution. As an example, Figure~\ref{figpre18} depicts projections on the plane $(x,y)$ of two trajectories $\gamma_{1,2}$, obtained as solutions of the system \eqref{(1)} with close initial conditions. Both trajectories approach uncertainty curve along the same guiding trajectory, but diverge after that -- $\gamma_1$ exits ${Q_2}(\xi )$ and follows the outgoing guiding trajectory in ${Q_1}(\xi )$, while $\gamma_2$ turns and goes back along the other guiding trajectory in ${Q_2}(\xi )$.

In deterministic systems with strongly entangled trajectories the probability of a specific outcome is determined by the fraction of phase volume occupied by corresponding initial conditions (formal definition can be found in \cite{Arnold} and \cite{Neishtadt_2DoF}). In order to find the probabilities of transitions to different outgoing trajectories in some point $(x_*,y_*)$ on $\Gamma_2$, two auxiliary parameters must be calculated first \citep{Neishtadt_2DoF, Artemyev}:
\begin{equation}\label{(23)}
\Theta_{1,2}=\int\limits_{-\infty}^{+\infty} {{{\left({\frac{{\partial W_{\max}^*}}{{\partial x}}\frac{\partial W}{\partial y} - \frac{\partial W_{\max }^*}{\partial y}\frac{{\partial W}}{\partial x}} \right)}_{\varphi_{1,2}^s(t,x_*,y_*,\xi )}}d\tau}.
\end{equation}
These parameters have a meaning of rates with which areas bounded by separatrices $(\varphi _1^s(\tau,{x_*},{y_*},\xi ),\Phi _1^s(\tau,{x_*},{y_*},\xi ))$ and $(\varphi _2^s(\tau,{x_*},{y_*},\xi ),\Phi _2^s(\tau,{x_*},{y_*},\xi ))$ in fast variables' phase space change. After substitution of specific potential function \eqref{(2)} into \eqref{(23)} and change of integration variable to $\varphi$ we obtain:
\begin{equation}\label{(24)}
{\Theta _1} = 2\int\limits_{\varphi _{\min }^s}^{\phiarch } {\frac{\sin (\varphi  - \phiarch )}{\sqrt {2\left(\xi  - W({x_*},{y_*},\varphi)\right)}}d\varphi },
\quad
{\Theta _2} = 2\int\limits_{\phiarch }^{\varphi _{\max }^s} {\frac{{\sin (\varphi  - \phiarch )}}{{\sqrt {2\left(\xi  - W({x_*},{y_*},\varphi)\right)} }}d\varphi }.
\end{equation}

Here $\phiarch$ is the coordinate $\varphi$ of the saddle point in the fast subsystem's phase portrait (it coincides with the value $\phiarch$ corresponding to point $(x_*,y_*)$ in~\eqref{(12)}); $\varphi _{\min }^s$ and $\varphi _{\max }^s$ are the minimal and the maximal values of $\varphi $ in homoclinic trajectories to the left and to the right of the saddle point respectively. Applying the substitution $\lambda  = \cot[{{(\varphi  - \phiarch )} / 2}]$ to \eqref{(24)} we obtain:
\begin{equation}\label{(25)}
{\Theta _{1,2}} = \frac{4}{{\sqrt A }}\int\limits_{{\lambda _{\min }}}^{{\lambda _{\max }}} {\frac{{\lambda \;d\lambda }}{{(1 + {\lambda ^2})\sqrt {(\lambda  - a)(\lambda  - b)} }}},
\end{equation}
where
\[
a = \frac{B-\sqrt {B^2 - A C}}{A},
\quad
b = \frac{B+\sqrt {B^2 - A C}}{A},
\]
\[
A = \xi  + 3\cos (2\phiarch),
\quad
B = 2\sin (2\phiarch),
\quad
C = \xi  - \cos (2\phiarch).
\]

It should be noted, that inequalities $A>0$ and $B^2>AC$ hold for all points of $\Gamma _2$.
Integration in \eqref{(25)} is carried out over the intervals $(-\infty,a)$ and $(b,+\infty)$ for $\Theta_1$ and $\Theta_1$ respectively. The result can be obtained by using Cauchy's residue theorem:
\[
\Theta_1= \frac{4}{{\sqrt A }}{\mathop{\rm Re}\nolimits} \left( {\frac{{L - i\pi }}{{\alpha \beta }}} \right),
\quad
\Theta_2=\frac{4}{\sqrt A}{\mathop{\rm Re}\nolimits} \left( {\frac{L}{{\alpha \beta }}} \right).
\]
Here
\[
L = \ln \left(\frac{b-a}{(\alpha-\beta)^2}\right),
\quad
\alpha=\sqrt{a+i},
\quad
\beta=\sqrt{b+i},
\]
and the branches of multifunctions are selected in such way, that $\operatorname{Im}(L)$, $\arg(\alpha )$, and $\arg(\beta ) \in (0,\pi)$.

Now we can write down the expressions for probabilities of different evolution scenarios for a phase point on $\Gamma_2(\xi)$. Let us denote the probability of point going to region $Q_1(\xi)$ as $P_0$, and probabilities corresponding to two trajectories going inside $Q_2(\xi)$ as $P_1$ and $P_2$. The resulting formulae \citep{Artemyev} take a form:
\begin{equation}\label{(26)}
P_0=1-P_1-P_2,
\quad
P_1= \max (\Theta_1,0)/{\Theta_\Sigma},
\quad
P_2= \max (\Theta_2,0)/{\Theta_\Sigma},
\end{equation}
where
\[
\Theta_\Sigma=\max(\Theta_1,0)+\max(\Theta_2,0)+\max(-\Theta_1-\Theta_2,0).
\]

\begin{figure}[H]
\center{
\includegraphics[width=122mm]{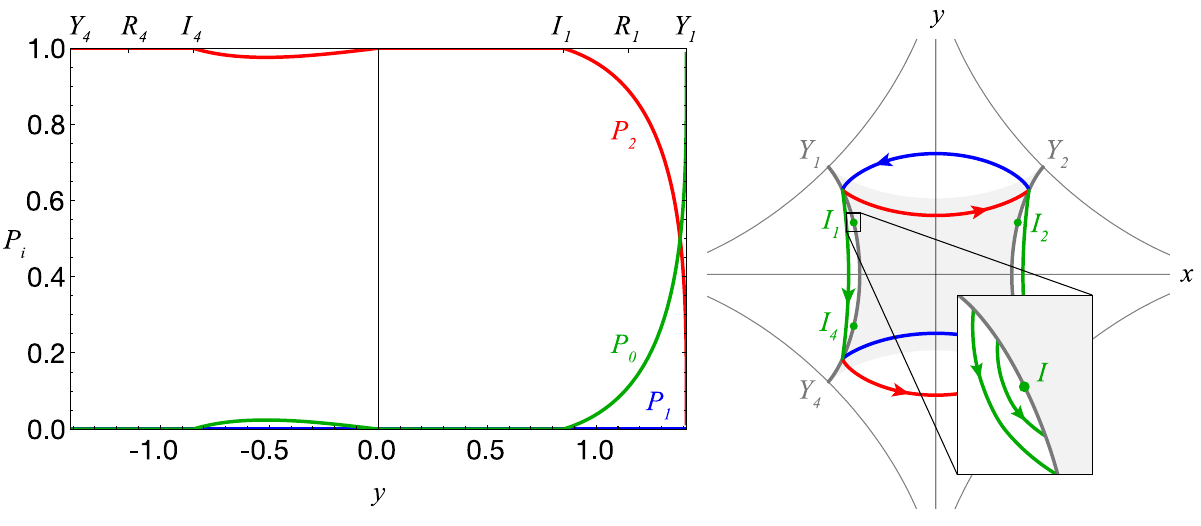}
}
\caption{Change of $P_i$ along $Y_4 Y_1$ segment of $\Gamma_2$ at $\xi=0$. The graph for $Y_2 Y_3$ can be reconstructed by the symmetry, changing the $y$ sign and swapping red and blue plots. In $I_1,\ldots,I_4$ the sum $\Theta_1+\Theta_2=0$, which results in $P_1$ plot sticking to $0$ on one side from these points}
\label{fig18}
\end{figure}

\begin{figure}[H]
\center{
\includegraphics[width=122mm]{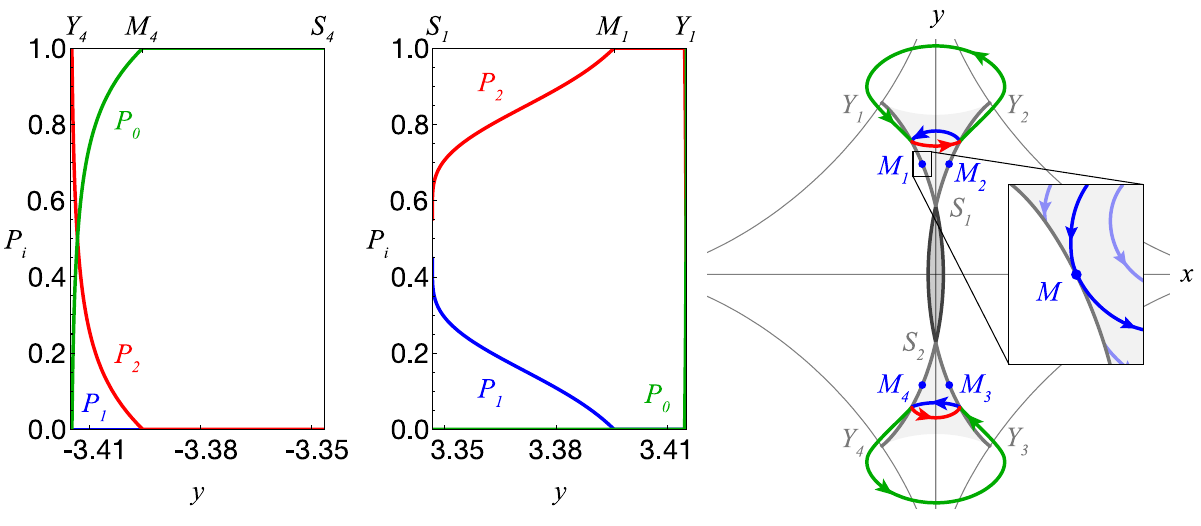}
}
\caption{Change of $P_i$ along segments $Y_4 S_2$ and $S_1 Y_1$ of $\Gamma_2$ at $\xi=2.4$. In $M_{1,2}$ parameter $\Theta_1=0$, and in $M_{3,4}$ parameter $\Theta_2=0$, which results in singularities of probabilities plots in these limiting points}
\label{fig19}
\end{figure}

In Figure~\ref{fig18} and Figure~\ref{fig19} change of probabilities \eqref{(26)} along $\Gamma_2$ is shown alongside with phase portraits at corresponding values of Hamiltonian and limiting points, which mark the change of sign in $\Theta_{1,2}$ or $\Theta_1+\Theta_2$.

\subsection{Adiabatic chaos}

\begin{figure}[ht]
\center{\includegraphics[width=80mm]{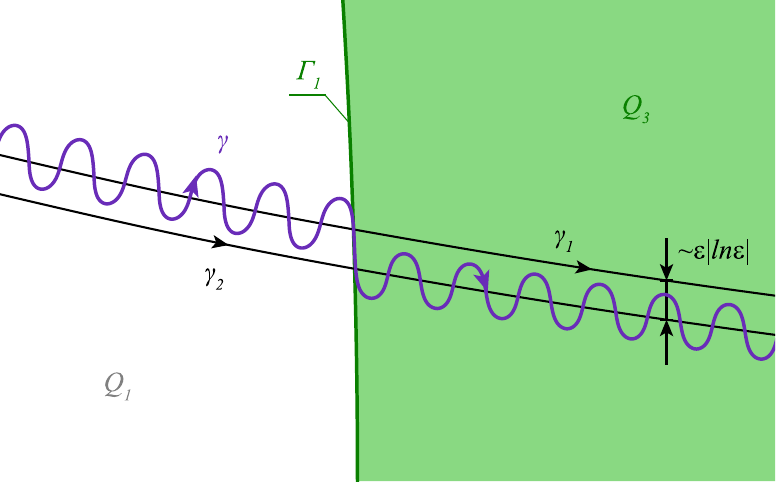}}
\caption{The jump of trajectory $\gamma$ from guiding trajectory $\gamma_1$ to $\gamma_2$ upon crossing the uncertainty curve}
\label{fig20}
\end{figure}

\emph{Adiabatic chaos} emerges due to non-applicability of adiabatic approximation near the uncertainty curve. As a result the projection of phase point $\boldsymbol{\zeta}(\tau)=(x(\tau),y(\tau))^T$ leaves the vicinity of uncertainty curve along the guiding trajectory, which slightly differs from the direct continuation of approach trajectory (Fig.~\ref{fig20}). The resulting offset between incoming and outgoing guiding trajectories can be treated as a quasi-random jump with order of magnitude $\varepsilon \left| \ln \varepsilon \right|$ \citep{Tennyson86, Neishtadt_2DoF, Neishtadt_kirkwood}.

\begin{figure}[ht]
\center{\includegraphics[width=80mm]{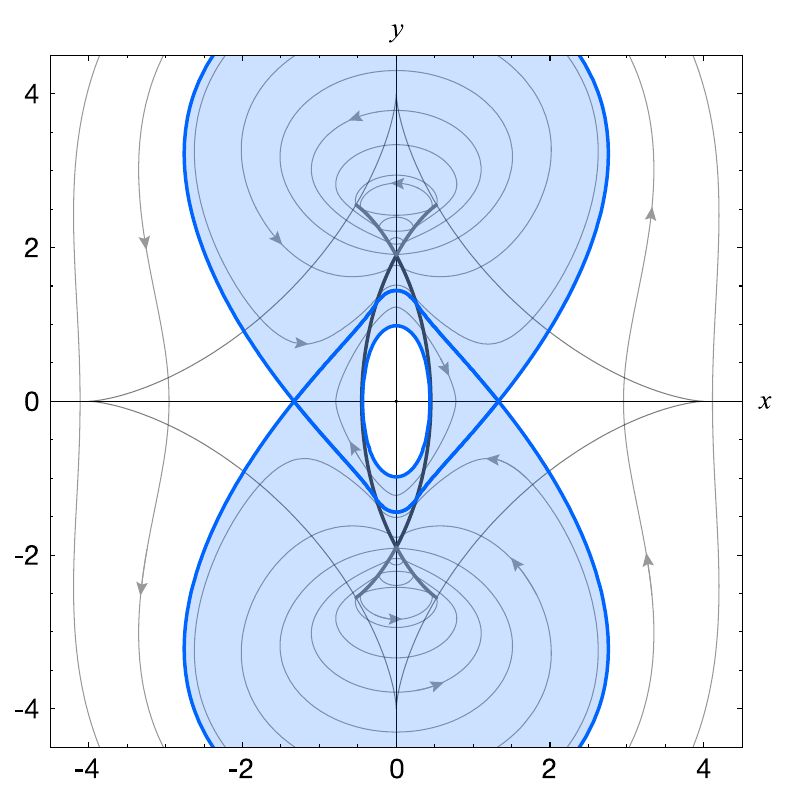}}
\caption{Adiabatic chaos region at $\xi=1.45$}
\label{fig21}
\end{figure}

As a result of persistent jumps any trajectory that crosses the uncertainty curve after a long time will fill a whole region of phase plane, which we will call an \emph{adiabatic chaos region}  (Fig.~\ref{fig21}). This region consists of points belonging to trajectories, which cross the uncertainty curve.

At $\xi \in (\xi_3, \xi_8)$ the uncertainty curve is crossed by the separatrices, which connect two saddle points. A phase point projection moving along a guiding trajectory close to separatrix, when crossing the uncertainty curve, may jump over the separatrix and begin to move along the other guiding trajectory belonging to completely different family. Thus the properties of long-term evolution are suddenly changed on a qualitative level. E.g., in Figure~\ref{fig21} the motion of a phase point projection $\boldsymbol{\zeta}(\tau)$ circling around the coordinate origin in the central part of adiabatic chaos region by crossing the separatrix may transform into circulation in opposite direction around one of two center points~\eqref{(22)}, which lie on $y$-axis in upper or lower half-plane. This event can be interpreted as a capture into Kozai-Lidov resonance and it is accompanied by decrease of average inclination value, about which the long-term oscillations occur.

\section{Numerical simulations}\label{Sec:num}

Construction of the discussed analytical model involved several assumptions that may seem loose. It is thus required to test whether the model can be applied to orbital dynamics of real life objects and these assumptions were not overly restrictive.

For this purpose we used Mercury integrator \citep{Mercury6} and carried out several numerical simulations of the Solar system composed of the Sun, the four giant planets, and about 700 known Kuiper belt objects (KBO) near $1:2$, $2:3$, and $3:4$ resonances with Neptune represented by test particles (masses of four inner planets were added to the Sun in order to facilitate the integration).
Total time of integration $15~Myr$ is one order of magnitude larger then the characteristic time $T_N/\varepsilon^2\approx1.6~Myr$ of slow variables evolution (given the orbital period of Neptune $T_N\approx160~y$ and Neptune/Sun mass ratio defining the small parameter $\varepsilon^2\sim 10^{-4}$).

\begin{figure}[H]
\center{
\includegraphics[width=122mm]{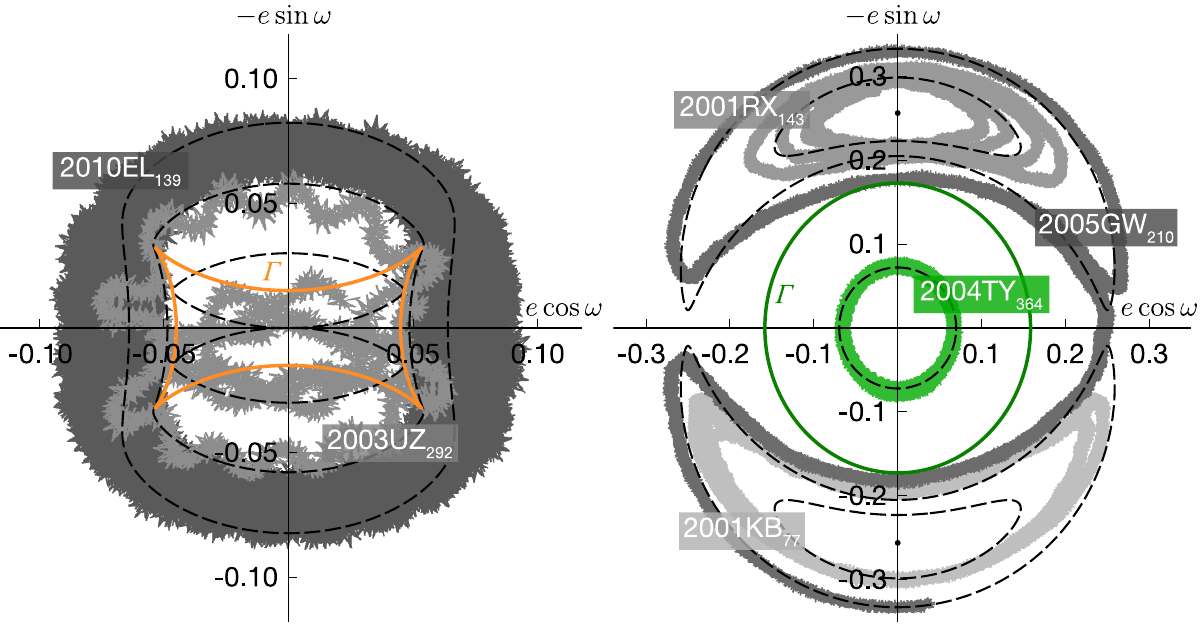}
}
\caption{Comparison of the Solar system's numerical integration with analytical model. Several Kuiper belt objects with Hamiltonian values close to $\xi\approx-0.5$ (left) and $\xi\approx23$ (right) are plotted on the plane of slow variables. Phase trajectories of the model plotted by dashed lines}
\label{num1}
\end{figure}

For interpretation of the simulation results we shall utilize a scaled version of previously used slow variables:
\[
x\propto e \cos \omega,\quad
y\propto - e \sin \omega.
\]
For exact relations between $(e,\omega)$ and $(x,y)$, as well as the expression for small parameter $\varepsilon$, see Appendix A.

Phase trajectories of several objects are plotted in Figure~\ref{num1}. Main sources of difference between analytical phase portraits and numerical ones are non-zero eccentricity of Neptune ($e'\approx0.01$), high eccentricity of KBOs (up to $0.3$ on the right side of Figure~\ref{num1}), and presence of other planets, which introduce additional disturbances distorting the phase plane. Nevertheless, it is clear that the overall topological structure of phase portraits is reproduced in analytical model.
Note, that in restricted three-body problem all solutions of \cite{FerrazMello84} on the same phase plane would be represented by concentric circles with $e=const$ and $\omega$ changing linearly with time.

The principal difference between our model and the one introduced by \cite{FerrazMello84} is the expansion of disturbing function past the first term in the Fourier series. Thus we can assert, that the region of the complete phase space, where the second term influences the dynamics in a substantial way, is significantly large. Indeed about $10\%$ of objects in our simulations deviated from circular trajectories in projection on the plane of slow variables. The rest correspond to very high or very low values of $\xi$, at which all trajectories in $Q_1$ and $Q_3$, as well as the curve $\Gamma$ separating them, in our model are likewise very close to circles.

\begin{figure}[H]
\center{
\includegraphics[width=122mm]{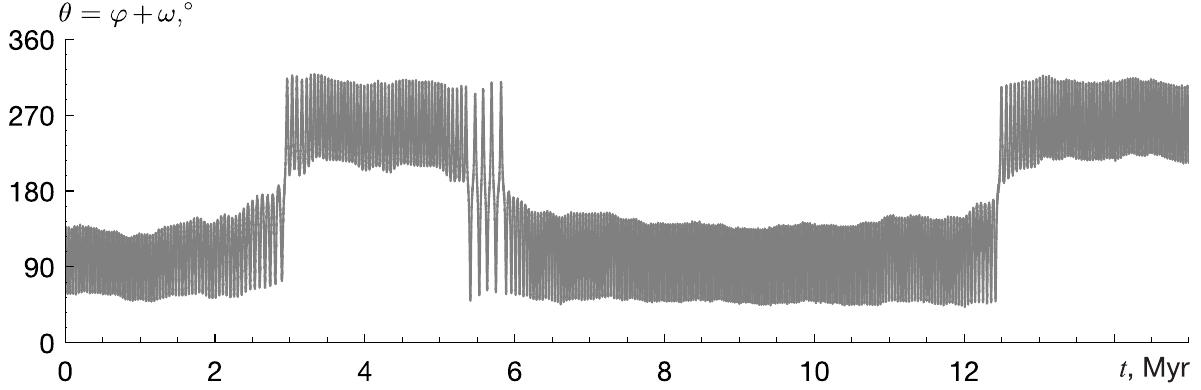}
}
\caption{Resonant angle of $2011UG_{411}$ vs time, demonstrating jumps between two different librating solutions in $Q_2$}
\label{num3}
\end{figure}

\begin{figure}[H]
\center{
\includegraphics[width=122mm]{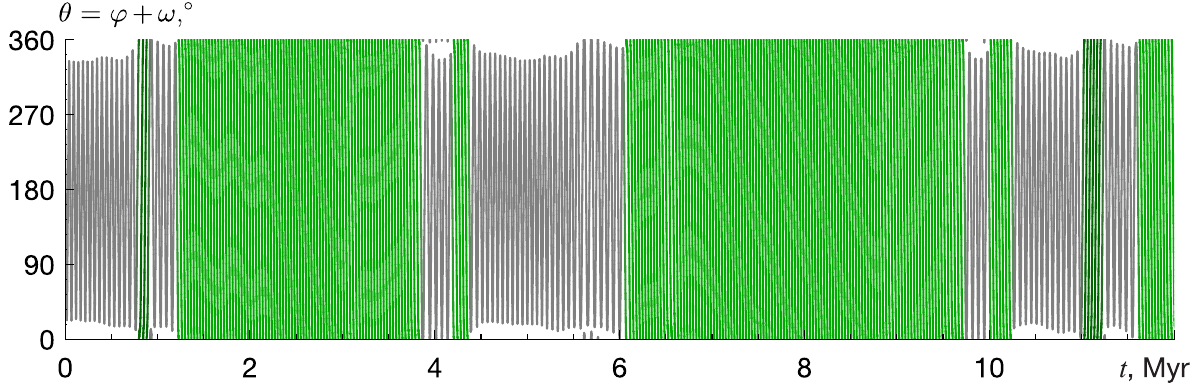}
}
\caption{Resonant angle of $2007JJ_{43}$ vs time. Intervals of circulation and libration colored green and gray respectively}
\label{num2}
\end{figure}

\begin{figure}[H]
\center{
\includegraphics[width=122mm]{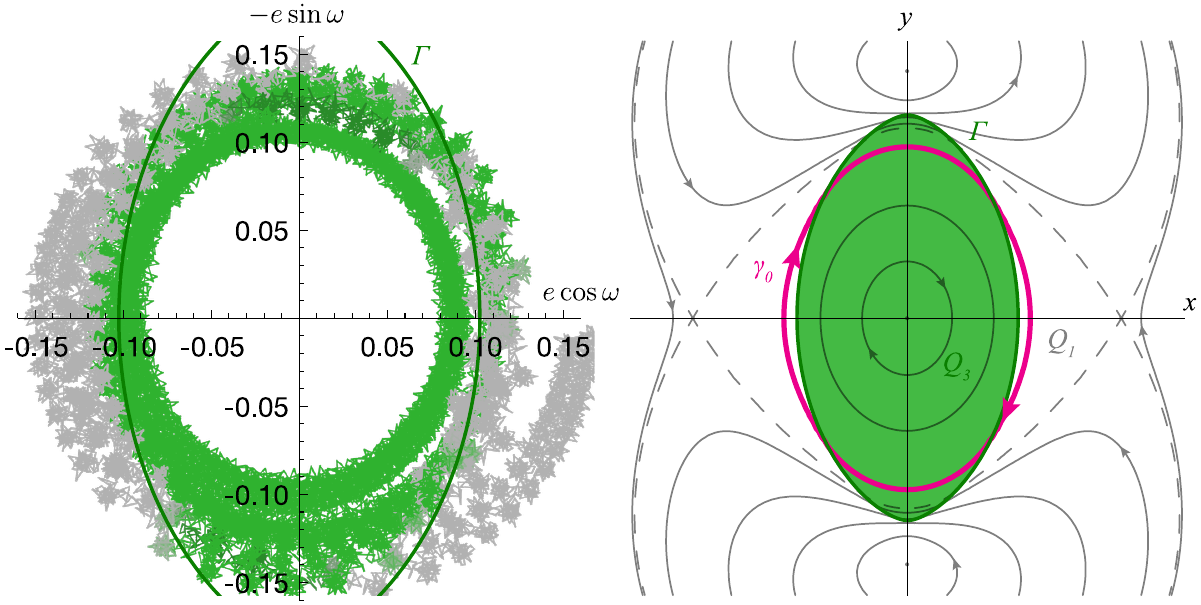}
}
\caption{Phase trajectory of $2007JJ_{43}$ alongside the analogous trajectory $\gamma_0$ on the analytical phase portrait. Intervals of resonant angle circulation and libration on the left panel are colored green and gray respectively. It is seen, that the green segments of the trajectory concentrate inside the region bound by $\Gamma$, while the gray ones mostly lie outside of it}
\label{num4}
\end{figure}

Some effects, that can be derived from the analytical model, are also observed in numerical simulations within the pool of selected KBOs. E.g. resonant angle of $2011UG_{411}$ shows jumps between two different librating regimes characteristic to motion in region $Q_2$ (Fig.~\ref{num3}), while $2007JJ_{43}$ demonstrates the intermittent behaviour (Fig.~\ref{num2}) with the resonance angle constantly switching between libration and circulation. Phase trajectory of $2007JJ_{43}$ on the plane of slow variables is presented in Figure~\ref{num4}, showing that changes in resonant angle behaviour are conditioned by secular trajectory crossing of the critical curve $\Gamma$. Similarly, analytical trajectory $\gamma_0$ goes between regions of libration and circulation of resonant angle (Fig.~\ref{num4}).
To compensate for angle $\omega$ precession on this kind of trajectories the modified resonant angle $\theta=\varphi+\omega$ was used in previous plots\footnote{The resonant angle $\theta$ is the ``standard'' one for studies of first-order MMR \citep{MurrayDermott}.
Our choice of the resonant angle $\varphi$ was determined by the desire to write down the Hamiltonian of the system in the form \eqref{(2)}, which seem to us most convenient for the analysis.}. Same as $\varphi$, resonant angle $\theta$ circulates in $Q_3$ and librates in $Q_1$ and $Q_2$.

\section{Conclusion}\label{conclusion}

In this paper, using the averaging technique, we study Hamiltonian system that approximately describes the dynamics of a three-body system in first-order MMR (within the restricted circular problem). Our model incorporates harmonics of the Fourier series expansion of disturbing function up to the second order. Thus it can accurately describe the dynamics in that part of the phase space where first two harmonics have comparable magnitudes, and where the well known integrable model of first-order MMR is not applicable. This is the region, from which chaos emerges.

The nonintegrability of our model  does not become an obstacle for a detailed analytical investigation of its properties. In particular, we have constructed bifurcation diagrams and phase portraits characterizing the long-term dynamics on different level sets defined by the system's Hamiltonian and obtained expressions for probabilities of quasi-random transitions between different phase trajectories.

The important question is the scope of the correctness of proposed model. We will try to answer it in subsequent studies, using Wisdom's approach to investigate several first-order MMRs without truncating the averaged disturbing function. Nevertheless, even now we can note that the secular evolution of some Kuiper belt objects qualitatively resembles what our simple model predicts.

We hope also that our investigation outlines an approach which can be applied for analysis of similar degeneracy of the averaged disturbing function in the case of other MMRs (e.g., \cite{Sidorenko_31}).

\section*{Appendix A: Constructing a model system, that reveals the origin of chaos in first-order MMR}

We shall confine ourselves to a case of exterior resonance $p:(p+1)$ in restricted three-body problem. The interior resonance $(p+1):p$ can be reduced to the same model using similar approach.
The distance between the two major bodies, i.e. a star and a planet, and the sum of their masses are taken here as units of length and mass. The unit of time is chosen such that the orbital period of major bodies' rotation about barycenter is equal to $2\pi$. The mass of the planet $\mu$ is considered to be a small parameter of the problem.

Equations of motion for minor body (asteroid) in canonical form:
\begin{equation}\label{B1}
\frac{d(L,G,H)}{dt} =  - \frac{\partial \mathcal{K}}{\partial (l,g,h)},\quad
\frac{d(l,g,h)}{dt} = \frac{\partial \mathcal{K}}{\partial (L,G,H)},
\end{equation}
where $L, G, H, l, g, h$ are the Delaunay variables \citep{MurrayDermott}. They can be expressed in terms of
Keplerian elements $a,e,i,\Omega, \omega$ as
\[
L = \sqrt {(1 - \mu )a},\quad
G = L\sqrt {1 - {e^2}},\quad
H = G\cos i,\quad
g = \omega ,\quad
h = \Omega.
\]
The last variable $l$ is the asteroid's mean anomaly.

Hamiltonian $\mathcal{K}$ in \eqref{B1} is
\begin{equation}\label{B2}
\mathcal{K} =  - \frac{{{{(1 - \mu )}^2}}}{{2{L^2}}} - \mu R(L,G,H,l,g,h - \lambda ').
\end{equation}
Here $R$ is disturbing function in restricted circular three-body problem. Mean anomaly $\lambda '$ appearing in \eqref{B2} depends linearly on time: $\lambda ' = t + {\lambda '_0}$. Therefore it is convenient to use variable $\tilde h = h - \lambda '$ instead of $h$, as it enables writing down the equations of motion in autonomous form as canonical equations with Hamiltonian
\[
 \tilde{ \mathcal{K}} = \mathcal{K}(L,G,H,l,g,\tilde h) - H.
\]

We introduce resonant angle $\bar\varphi$ using the canonical transformation $(L,G,H,l,g,\tilde h) \to ({P_\varphi },{P_g},{P_h},\bar\varphi ,\bar g,\bar h)$ defined by generating function
\[
S = (p + 1){P_\varphi }l + \left[ {{P_h} + p\left( {{P_\varphi } - P_\varphi ^*} \right)} \right]\tilde h + \left[ {{P_g} + (p + 1)\left( {{P_\varphi } - P_\varphi ^*} \right)} \right]g,
\]
where $P_\varphi ^* = L_*/(p+1)$, while $L_*=\sqrt[3]{(p+1)/p}$ is the value of $L$ corresponding to exact $p:(p+1)$ MMR in unperturbed problem ($\mu=0$). The new variables are related to the old ones as follows:
\begin{gather*}
L = \frac{{\partial S}}{{\partial l}} = (p + 1){P_\varphi },\quad
G = \frac{{\partial S}}{{\partial g}} = {P_g} + (p + 1)\left( {{P_\varphi } - P_\varphi ^*} \right),\\
H = \frac{{\partial S}}{{\partial \tilde h}} = {P_h} + p\left( {{P_\varphi } - P_\varphi ^*} \right),\\
\bar\varphi  = \frac{{\partial S}}{{\partial {P_\varphi }}} = (p + 1)l + p\tilde h + (p + 1)g,\quad
\bar g = \frac{{\partial S}}{{\partial {P_g}}} = g,\quad
\bar h = \frac{{\partial S}}{{\partial {P_h}}} = \tilde h.
\end{gather*}
The resonant angle can also be expressed in traditional form
\[
\bar\varphi  = (p + 1)\lambda  - p\lambda' - \Omega.
\]
New Hamiltonian
\[
\tilde {\mathcal K}=-\frac{(1-\mu)^2}{2(p+1)^2 P_\varphi^2}-\left[P_h+p(P_\varphi-P_\varphi^*)\right] - \mu R.
\]

The resonant case we are interested in corresponds to region $\mathcal{R}$ of phase space, selected by the condition
\[
\left| {pn' - (p + 1)n} \right| \lesssim {\mu ^{1/2}}.
\]
Here $n'=1$ and $n$ are mean motions of planet and asteroid respectively. It is also true in $\mathcal{R}$ that
\begin{equation}\label{resonancecondition}
\left| {{P_\varphi } - P_\varphi ^*} \right| \lesssim {\mu ^{1/2}}.
\end{equation}
In the resonant case, i.e. when the previous inequation holds, variables can be divided into \emph{fast}, \emph{semi-fast} and \emph{slow}. Fast and semi-fast variables in $\mathcal{R}$ are $\bar h$ and $\bar\varphi$ respectively:
\[
\frac{{d\bar h}}{{dt}} \sim 1,\quad
\frac{{d\bar\varphi }}{{dt}} \sim {\mu ^{1/2}}.
\]
Slow variables, which vary with a rate of order $\mu$, are $P_\varphi$, $P_g$, $P_h$, and $\bar g$.

To study secular effects the averaging over the fast variable $\bar h$ is performed, which results in equations of motion taking a canonical form with Hamiltonian
\begin{multline*}
{\bar{ \mathcal K}}({P_\varphi },{P_g},{P_h},\bar\varphi ,\bar g) =\\
=\frac{1}{{2\pi (p + 1)}}\int\limits_0^{2\pi (p + 1)} {\tilde {\mathcal K}(L({P_\varphi }),G({P_g},{P_\varphi }),H({P_h},{P_\varphi }),l(\bar\varphi ,\bar g,\bar h),\bar g,\bar h)} \,d\bar h,
\end{multline*}
where
\[
l(\bar\varphi ,\bar g,\bar h) = \frac{1}{{k + 1}}\left[ {\bar\varphi  - (k + 1)\bar g - k\bar h} \right].
\]
After such averaging the fast variable $\bar{h}$ vanishes, and the term ``fast'' is exempted. Thus in the rest of the paper we adopt name \emph{fast} for denoting variables, which vary with the rate $\mu^{1/2}$, instead of referring to them as semi-fast, when the distinction of three different time scales was needed.

Moreover, because there is no longer $\bar h$ in the Hamiltonian $\bar{ \mathcal K}$, the conjugate momentum $P_h$ is a constant in considered approximation and can be treated as a parameter of the problem. Thus $\bar{ \mathcal K}$ is the Hamiltonian of a system with two degrees of freedom.
Further instead of $P_h$ we shall use parameter
\[
\sigma  = \sqrt {1 - \frac{P_h^2}{L_*^2}}.
\]
Inequation $e\le\sigma$ defines region $\mathcal S$ in phase space, to which the motion of the system is bound by the Kozai-Lidov integral \citep{Sidorenko_QS}.

Next standard step in analysis of system's dynamics in resonant region $\mathcal R$ is the scaling transformation \citep{AKN_eng}:
\begin{equation}\label{B3}
\bar\tau  = \bar\varepsilon t,\quad
\bar\Phi  = (P_\varphi ^* - {P_\varphi })/\bar\varepsilon,
\end{equation}
where $\bar\varepsilon  = {\mu ^{1/2}}$ is a new small parameter. Using variables \eqref{B3}, the equations of motion can be rewritten in a form of slow-fast system without loss of accuracy:
\begin{equation}\label{B4}
\begin{gathered}
\frac{d\bar\varphi}{d\bar\tau} = \chi \bar\Phi,\quad
\frac{d\bar\Phi}{d\bar\tau} =  - \frac{{\partial \bar{W}}}{{\partial \bar\varphi }},\\
\frac{d P_g}{d\bar\tau} = \bar\varepsilon \frac{{\partial \bar{W}}}{{\partial \bar g}},\quad
\frac{d\bar g}{d\bar\tau} =  - \bar\varepsilon \frac{{\partial \bar{W}}}{{\partial {P_g}}}.
\end{gathered}
\end{equation}
Here
\[
\begin{aligned}
&\chi  = 3{p^{4/3}}{\left(p + 1\right)^{2/3}},
\\
\bar {W}(\bar \varphi ,\bar g,{P_g};\sigma ) = \frac{1}{{2\pi (k + 1)}}&\int\limits_0^{2\pi (k + 1)} {R\left({L_*},{P_g},{L_*}\sqrt {1 - {\sigma ^2}} ,l(\bar \varphi ,\bar g,\bar h),\bar g,\bar h \right)} \,d\bar h.
\end{aligned}
\]
For $\sigma\ll 1$ the approximate expression for $\bar {W}(\bar \varphi ,\bar g,{P_g};\sigma )$ can be obtained as a series expansion \citep{MurrayDermott}:
\begin{multline}\label{B5}
\bar{W}(\bar \varphi ,\bar g,{P_g};\sigma ) \approx {W_0} + {W_1}e\cos (\bar \varphi  - \bar g) + \\
 + {e^2}\left[ {{W_{10}} + {W_{11}}\cos 2(\bar \varphi  - \bar g)} \right] + {i^2}\left[ {W_{20} + {W_{21}}\cos 2\bar \varphi } \right],
\end{multline}
where
\begin{equation}\label{B6}
  e^2\approx1-\frac{P_g^2}{L_*^2},\quad
   i^2\approx 2\left(1-\frac{P_h}{L_* \sqrt{1-e^2}}\right)\approx \sigma^2-e^2.
\end{equation}
Expressions \eqref{B6} are obtained taking into account resonance condition \eqref{resonancecondition}, and that the values $e$ and $i$ are limited by $\sigma\ll 1$, leading to $e, i\ll 1$.

Coefficients $W_0$, $W_1$, $W_{10}$, $W_{11}$, $W_{20}$, $W_{21}$ in \eqref{B5} are calculated as follows:
\begin{equation}\label{BW}
\begin{array}{l}
W_0 = \frac{\alpha }{2}b_{1/2}^{(0)},\quad
W_1 = \frac{\alpha }{2}\left[ {(2p + 1)b_{1/2}^{(p)} + \alpha \frac{{db_{1/2}^{(p)}}}{{d\alpha }}}\right] - \frac{\delta_{1p}}{2\alpha},\\
W_{10} = \frac{\alpha ^2}{8}\left({2\frac{{db_{1/2}^{(0)}}}{{d\alpha }} + \alpha \frac{{{d^2}b_{1/2}^{(0)}}}{{d{\alpha ^2}}}} \right),\\
W_{11} = \frac{\alpha }{8}\left( {\left[ {2 - 14(p + 1) + 16{{(p + 1)}^2}} \right]b_{1/2}^{(2p)} + \alpha \left[ {8(p + 1) - 2} \right]\frac{{db_{1/2}^{(2p)}}}{{d\alpha }} + {\alpha ^2}\frac{{{d^2}b_{1/2}^{(2p)}}}{d\alpha^2}} \right),\\
W_{20}= - W_{10},\quad
{W_{21}} = \frac{{{\alpha ^2}}}{8}b_{3/2}^{(2p + 1)}.
\end{array}
\end{equation}
Here $\alpha  = {(p/(p + 1))^{2/3}}$, $\delta_{mn}$ is the Kronecker delta, and $b_{1/2}^{(n)}(\alpha )$, $b_{3/2}^{(n)}(\alpha )$ are Laplace coefficients. Numerical values of coefficients \eqref{BW} for several resonances are gathered in Table~\ref{tbl:W}.

\begin{table}[H]
\caption{Numerical values of coefficients in series expansion of averaged disturbing function}
\label{tbl:W}
\begin{center}
\begin{tabular}{lccccc}
\toprule
            & $p=1$     & $p=2$     & $p=3$     & $p=4$     & $p=5$\\
\midrule
 $W_0$      & $0.7120$  & $0.9381$  & $1.0767$  & $1.1760$  & $1.2531$\\
\midrule
 $W_1$      & $0.2699$  & $1.8957$  & $2.7103$  & $3.5192$  & $4.3257$\\
\midrule
 $W_{10}$   & $0.2442$  & $0.8798$  & $1.8815$  & $3.2449$  & $4.9686$\\
\midrule
 $W_{11}$   & $2.2640$  & $6.3052$  & $12.260$  & $20.133$  & $29.923$\\
\midrule
 $W_{21}$   & $0.1291$  & $0.4375$  & $0.9160$  & $1.5640$  & $2.3814$\\
\bottomrule
\end{tabular}
\end{center}
\end{table}

Out of the whole region $\mathcal S$ we are interested in the part with small eccentricities. Thus we further assume $e/\sigma\lesssim \sigma$, which leads to $e\lesssim \sigma^2$. Taking into account \eqref{B6} we can write down expression for averaged disturbing function up to the terms of order~$\sigma^2$:
\[
\bar{W}(\bar \varphi ,\bar g,{P_g};\sigma ) \approx {W_0} + {W_1}e\cos (\bar \varphi  - \bar g) + {\sigma ^2}\left[ {{W_{20}} + {W_{21}}\cos 2\bar \varphi } \right].
\]
By introducing the variables
\[
\bar x = \sqrt {2({L_*} - {P_g})} \cos \bar g,\quad
\bar y =  - \sqrt {2({L_*} - {P_g})} \sin \bar g,
\]	
the equations of motion \eqref{B4} can be reduced to a Hamiltonian form
\[
\begin{gathered}
\frac{{d\bar\varphi }}{{d\bar \tau }} = \frac{\partial \bar {\Xi}}{{\partial \bar \Phi }},\quad
\frac{{d\bar\Phi }}{{d\bar \tau }} =  - \frac{\partial \bar {\Xi}}{{\partial \bar \varphi }},\\
\frac{{d\bar x}}{{d\bar \tau }} = \bar\varepsilon \frac{\partial \bar{\Xi}}{{\partial \bar y}},\quad
\frac{{d\bar y}}{{d\bar \tau }} =  - \bar\varepsilon \frac{\partial \bar{\Xi}}{{\partial \bar x}}
\end{gathered}
\]
with the Hamiltonian
\[
\bar {\Xi}  = \frac{{\chi {\bar{\Phi} ^2}}}{2} + \frac{{{W_1}}}{{\sqrt {{L_*}} }}\cos \bar\varphi  \cdot \bar x - \frac{{{W_1}}}{{\sqrt {{L_*}} }}\sin \bar \varphi  \cdot \bar y + {\sigma ^2}{W_{21}}\cos 2\bar\varphi.
\]
The final rescaling of variables
\begin{equation}\label{B7}
\begin{gathered}
\varphi=  - \bar{\varphi},\quad
\Phi= - \sqrt {\frac{\chi }{{{\sigma ^2}{W_{21}}}}} \bar{\Phi},\\
x= \frac{{{W_1}}}{{{\sigma ^2}{W_{21}}\sqrt {{L_*}} }}\bar{x},\quad
y= \frac{{{W_1}}}{{{\sigma ^2}{W_{21}}\sqrt {{L_*}} }}\bar{y},\\
\tau= \sigma \sqrt {\chi {W_{21}}} \bar{\tau},\quad
\varepsilon= \frac{{{W_1}^2}}{{{{\chi^{1/2} {L_*}}}{\sigma ^3}W_{21}^{3/2}}}\bar{\varepsilon}
\end{gathered}
\end{equation}
results in the model system \eqref{(1)}--\eqref{(2)}.


\section*{Appendix B: Analytical expressions for integrals, emerging during the averaging over fast subsystem's period}

Right-hand side parts of evolution equations \eqref{(4)}, as well as adiabatic invariant formula \eqref{(21)}, can be expressed in terms of integrals
\[
I_{k,r} = \int\limits_{\lambda_*}^{\lambda^*} \frac{{\lambda ^r}d\lambda}{{({\lambda ^2} + 1)^k}\sqrt{\pm (\lambda-a_1)(\lambda-a_2)(\lambda-a_3)(\lambda-a_4)}},
\]
where $k = 0,1,2$; $r = 0,1$, and integration limits $\lambda_*$, $\lambda^*$ can be either real numbers or $\pm \infty$. These integrals can be reduced to the linear combinations of elliptic integrals of the first, the second and the third kind:
	
\begin{equation}\label{(A1)}
\begin{array}{*{20}{l}}
I_{0,0} = c_{0,0}K(k),\\
I_{1,0} = c_{1,1}K(k) + c_{1,3}\Pi \left( {h,k} \right) + {{\bar c}_{1,3}}\Pi \left( {\bar h,k} \right),\\
I_{1,1} = g_{1,1}K(k) + g_{1,3}\Pi \left( {h,k} \right) + {{\bar g}_{1,3}}\Pi \left( {\bar h,k} \right),\\
I_{2,0} = \frac{1}{2}{I_{1,0}} + {c_{2,1}}K(k) + {c_{2,2}}E(k) + {c_{2,3}}\Pi \left( {h,k} \right) + {{\bar c}_{2,3}}\Pi \left( {\bar h,k} \right),\\
I_{2,1} = {g_{2,1}}K(k) + g_{2,2}E(k) + {g_{2,3}}\Pi \left( {h,k} \right) + {{\bar g}_{2,3}}\Pi \left( {\bar h,k} \right).
\end{array}
\end{equation}

Further the formulae for coefficients $c_{m,l}$, $g_{m,l}$, moduli $k$ and parameters $h$ are gathered for all necessary cases.

\subsection*{The case ${a_j} \in {\mathbb{R}^1}$ $(j = \overline {1,4} )$}
We will further assume, that $a_j$ are numbered in ascending order:
\[
a_1<a_2<a_3<a_4.
\]
Four different instances should be considered:
\begin{enumerate}[label=\Alph*.]
\item Integration over $(a_1,a_2)$;
\item Integration over $(a_2,a_3)$;
\item Integration over $(a_3,a_4)$;
\item Integration over $(-\infty,a_1)\bigcup(a_4,+\infty)$.
\end{enumerate}

\subsubsection*{Instance A}
Here $\lambda _*=a_1$ and $\lambda^*=a_2$. We shall also use the following auxiliary parameters:
\[
{A_0} = \frac{2}{{\sqrt {({a_4} - {a_2})({a_3} - {a_1})} }},\quad
{C_0} = \frac{1}{{{a_2} - i}},\quad
{\alpha^2} = \frac{{{a_2} - {a_1}}}{{{a_3} - {a_1}}},\quad
\alpha_1^2 = \frac{{({a_2} - {a_1})(i - {a_3})}}{{({a_3} - {a_1})(i - {a_2})}}.
\]
The value of elliptic integrals' modulus in \eqref{(A1)}:
\[
k=\sqrt{\frac{(a_4-a_3)(a_2-a_1)}{(a_4-a_2)(a_3-a_1)}},
\]
and parameter $h = \alpha_1^2$.

In order to find coefficients in \eqref{(A1)}, we considered such linear combinations of ${I_{k,r}}$, that are reduced to integral (253.39) from \citep{ByrdFriedman}:
\begin{equation}\label{(A2)}
\begin{array}{*{20}{l}}
I_{1,1} + i I_{1,0} = \int\limits_{a_1}^{a_2} {\frac{{d\lambda}}{{(\lambda-i)\sqrt { - (\lambda-a_1)(\lambda-a_2)(\lambda-a_3)(\lambda-a_4)} }}} ,\\
I_{2,0} - i I_{2,1} - \frac{1}{2}{I_{1,0}} =  - \frac{1}{2}\int\limits_{{a_1}}^{{a_2}} {\frac{{d\lambda }}{{{{(\lambda  - i)}^2}\sqrt { - (\lambda-a_1)(\lambda-a_2)(\lambda -a_3)(\lambda-a_4)} }}}.
\end{array}
\end{equation}
After some simple calculations, one can find:
\begin{equation}\label{(A3)}
\begin{array}{*{20}{l}}
g_{1,1} = \operatorname{Re} C_1,\quad
c_{1,1} = \operatorname{Im} C_1,\quad
C_1 = \frac{{{A_0}{C_0}{\alpha ^2}}}{{\alpha _1^2}},\\
g_{1,3} = \frac{{{A_0}{C_0}}}{2}\left( {1 - \frac{{{\alpha ^2}}}{{\alpha _1^2}}} \right),\quad
c_{1,3} = - i g_{1,3},\\
c_{2,1} = - \operatorname{Re}{C_2},\quad
g_{2,1} = \operatorname{Im}{C_2},\quad
C_2 = \frac{{{A_0}C_0^2}}{{2\alpha _1^4}}\left[ {{\alpha ^4} + \frac{{{{(\alpha _1^2 - {\alpha ^2})}^2}}}{{2(\alpha _1^2 - 1)}}} \right],\\
c_{2,2} =  - \operatorname{Re}{C_3},\quad
g_{2,2} = \operatorname{Im}{C_3},\quad
C_3 = \frac{{{A_0}C_0^2{{(\alpha _1^2 - {\alpha ^2})}^2}}}{{4\alpha _1^2(\alpha _1^2 - 1)({k^2} - \alpha _1^2)}},\\
c_{2,3} = - \frac{{{A_0}C_0^2(\alpha _1^2 - {\alpha ^2})}}{{4\alpha _1^4}}\left[ {2{\alpha ^2} + \frac{{(2\alpha _1^2{k^2} + 2\alpha _1^2 - \alpha _1^4 - 3{k^2})(\alpha _1^2 - {\alpha ^2})}}{{2(\alpha _1^2 - 1)({k^2} - \alpha _1^2)}}} \right],\\
g_{2,3} = i{c_{2,3}}.
\end{array}
\end{equation}
Using (253.00) from \cite{ByrdFriedman} we also obtain:
\begin{equation}\label{(A4)}
c_{0,0} = A_0.
\end{equation}

\subsubsection*{Instance B}
For integrals over the interval $(a_2,a_3)$ the only difference is the parameters ${\alpha ^2}$, $\alpha _1^2$ and modulus $k$. Thus in \eqref{(A3)} the values
\[
\alpha^2= \frac{a_3 - a_2}{a_3-a_1},\quad
\alpha _1^2 = \frac{(a_3-a_2)(i-a_1)}{(a_3-a_1)(i-a_2)},\quad
k = \sqrt {\frac{(a_3 - a_2)(a_4-a_1)}{(a_4-a_2)(a_3-a_1)}}
\]
should be used.

\subsubsection*{Instance C}
For integrals over the interval $(a_3,a_4)$ in \eqref{(A3)}:
\[
C_0= \frac{1}{a_3-i},\quad
\alpha^2= \frac{a_4-a_3}{a_4-a_2},\quad
\alpha_1^2 = \frac{(a_4-a_3)(i-a_2)}{(a_4-a_2)(i-a_3)}.
\]
The modulus $k$ and parameter $A_0$ are the same as for interval $(a_1,a_2)$.

\emph{Note:} The equality of $k$ and $c_{0,0}$ in instances $A$ and $C$ means, that when the potential \eqref{(2)} has two minima, the periods of librations $T$ about them on the same energy level are equal. This also holds true for local minima corresponding to instances $B$ and $D$, as the substitution $\lambda=\tan[(\varphi-\tilde \varphi)/2]$ always allows to get rid of semi-infinite intervals of integration. Moreover, the averaged values of $\cos$ in \eqref{(17)} are also the same for librations around two local minima. The averaged values of $\sin$ in \eqref{(17)} for librations about two local minima differ by $\pi$, and values of adiabatic invariant~\eqref{(21)} differ by $y/\sqrt{2}$.

\subsubsection*{Instance D}
For integrals over two semi-infinite intervals the values of parameters in \eqref{(A3)} are
\[
C_0=\frac{1}{a_4-i},\quad
\alpha^2 = \frac{a_4-a_1}{a_3-a_1},\quad
\alpha_1^2 = \frac{(a_4-a_1)(i-a_3)}{(a_3-a_1)(i-a_4)}.
\]
The modulus $k$ is the same as for integration over $({a_2},{a_3})$, and $A_0$ is the same as for $({a_1},{a_2})$.

\subsection*{The case $a_1, a_2 \in {\mathbb{R}^1}$ ($a_1<a_2$), $a_3, a_4 \in {\mathbb{C}^1}$ ($a_3= \bar a_4$)}
Here there are only two instances:
\begin{enumerate}[label=\Alph*.]
\item Integration over $(a_1,a_2)$;
\item Integration over $(-\infty,a_1)\bigcup(a_2,+\infty)$.
\end{enumerate}

\subsubsection*{Instance A}
Here the following auxiliary parameters in expressions for $c_{m,l}$ and $g_{m,l}$ will be used:
\[
\alpha  = \frac{(a_1 A_2-a_2 A_1)-(A_2-A_1)i}{(a_1 A_2+a_2 A_1)-(A_2+A_1)i},\quad
\alpha_1= \frac{A_2-A_1}{A_2+A_1},
\]
\[
C_0= \frac{A_1+A_2}{(A_2 a_1-A_1 a_2)-(A_2-A_1)i},
\]
where
\[
\begin{array}{*{20}{l}}
{A_1} = \sqrt {(a_1-a_0)^2 + b_0^2},\quad
{A_2} = \sqrt {(a_2-a_0)^2 + b_0^2},\quad
{A_0} = 1/\sqrt{A_1 A_2},\\
{a_0} = \operatorname{Re} a_3 = \operatorname{Re} a_4,\quad
{b_0} = \operatorname{Im} a_3 = -\operatorname{Im} a_4.
\end{array}
\]
Using \eqref{(A2)} and formulae (259.04), (341.01)--(341.04)\footnote{It should be noted, that (259.04) in \citep{ByrdFriedman} contain a typo -- the multiplier $g$ (in authors' notation) is missing.} from \citep{ByrdFriedman}, we obtain
\begin{equation}\label{(A5)}
\begin{array}{*{20}{l}}
c_{0,0} = 2{A_0},\quad
g_{1,1} = \operatorname{Re} C_1,\quad
c_{1,1} = \operatorname{Im} C_1,\quad
C_1 = 2 A_0 C_0 \alpha_1,\\
c_{1,3} = - \frac{i(\alpha  - {\alpha _1})}{{(1 - {\alpha ^2})}}{A_0}{C_0},\quad
g_{1,3} = i c_{1,3},\\
c_{2,1} =  - \operatorname{Re} C_2,\quad
g_{2,1} = \operatorname{Im} C_2,\quad
C_2 = {A_0}C_0^2\left[ {\alpha _1^2 + \frac{{{{(\alpha  - {\alpha _1})}^2}}}{{{\alpha ^2} - 1}}} \right],\\
c_{2,2} =  -\operatorname{Re} C_3,\quad
g_{2,2} = \operatorname{Im} C_3,\quad
C_3 = \frac{{A_0 C_0^2 \alpha^2{{(\alpha-\alpha_1)}^2}}}{{(1-\alpha^2)(k^2 + {\alpha ^2}{{k'}^2})}},\\
c_{2,3} = \frac{{{A_0}C_0^2}}{2}\frac{{(\alpha  - {\alpha _1})}}{{({\alpha ^2} - 1)}}\left[ {2{\alpha _1} + \frac{{(\alpha  - {\alpha _1})\left[ {{\alpha ^2}(2k - 1) - 2{k^2}} \right]}}{{({\alpha ^2} - 1)({k^2} + {\alpha ^2}{{k'}^2})}}} \right],\quad
{g_{2,3}} = i{c_{2,3}}.
\end{array}
\end{equation}
Modulus and parameter of elliptic integrals are calculated as follows:
\[
k = \sqrt {\frac{{{{({a_2} - {a_1})}^2} - {{({A_2} - {A_1})}^2}}}{4 A_1 A_2}},\quad
h = \frac{{{\alpha ^2}}}{{{\alpha ^2} - 1}}.
\]

\subsubsection*{Instance B}
Expressions \eqref{(A5)} stay the same. Auxiliary parameters are
\[
{\alpha  = \frac{{({a_2}{A_1} + {a_1}{A_2}) - ({A_2} + {A_1})i}}{{({a_2}{A_1} - {a_1}{A_2}) + ({A_2} - {A_1})i}},\quad
{\alpha _1} = \frac{{{A_2} + {A_1}}}{{{A_1} - {A_2}}},}
\]
\[
C_0= \frac{{{A_1} - {A_2}}}{{({A_2}{a_1} + {A_1}{a_2}) - ({A_2} + {A_1})i}}.
\]
The modulus $k$ of elliptic integrals is also different:
\[
k = \sqrt {\frac{{{{({A_2} + {A_1})}^2} - {{({a_2} - {a_1})}^2}}}{{4{A_1}{A_2}}}}.
\]

\subsection*{The case ${a_j} \in {\mathbb{C}^1}$ $(j =\overline{1,4})$}
Let us denote real parts of roots $a_j$ as $p_{1,2}$, and imaginary parts as $b_{1,2}$:
\[
a_{1,2}=p_1\pm b_1,\quad
a_{3,4}=p_2\pm b_2.
\]
Without loss of generality let us assume, that ${p_1} < {p_2}$, ${b_1} > 0$, and ${b_2} > 0$ (if ${p_1} = {p_2}$, then also ${b_1} < {b_2}$). In expressions for ${c_{m,l}}$ and ${g_{m,l}}$ the following auxiliary parameters will be used
\[
{A_1} = \sqrt {{{({p_2} - {p_1})}^2} + {{({b_2} - {b_1})}^2}},\quad
{A_2} = \sqrt {{{({p_2} - {p_1})}^2} + {{({b_2} + {b_1})}^2}},\quad
{A_0} = 2/({A_1} + {A_2}),
\]
\[
{g_1} = \sqrt {\frac{{4b_1^2 - {{({A_2} - {A_1})}^2}}}{{{{({A_2} + {A_1})}^2} - 4b_1^2}}},\quad
\alpha = \frac{{{b_1} + {g_1}({p_1} - i)}}{{{p_1} - {b_1}{g_1} - i}},\quad
{C_0} = \frac{1}{{{b_1} + {g_1}({p_1} - i)}}.
\]
Similar to previous case we find:
\[\begin{array}{*{20}{l}}
c_{0,0} = 2 A_0,\quad
c_{1,1} = \operatorname{Im} C_1,\quad
g_{1,1} = \operatorname{Re} C_1,\quad
C_1 = \frac{2\alpha (1 + g_1 \alpha )}{1 + \alpha^2} A_0 C_0,\\
g_{1,3} = \frac{\alpha^2(\alpha-g_1)A_0 C_0}{1+\alpha^2},\quad
c_{1,3} = -i g_{1,3},\quad
c_{2,1} = -\operatorname{Re} C_2,\quad
g_{2,1} = \operatorname{Im} C_2,\\
C_2 = A_0 C_0^2 \left[g_1^2 + \frac{{2{g_1}(\alpha  - {g_1})}}{{1 + {\alpha ^2}}} + \frac{{{{(\alpha  - {g_1})}^2}}}{{(1 + {\alpha ^2})({\alpha ^2} + {{k'}^2})}}\left( {\frac{{2{{k'}^2} + 2{\alpha ^2} - {\alpha ^2}{k^2}}}{{1 + {\alpha ^2}}} - {{k'}^2}} \right) \right],\\
c_{2,2} = \operatorname{Re} C_3,\quad
g_{2,2} =  - \operatorname{Im} C_3,\quad
C_3 = \frac{{{A_0}C_0^2{{(\alpha  - {g_1})}^2}{\alpha ^2}}}{{({\alpha ^2} + 1)({\alpha ^2} + {{k'}^2})}},\\
c_{2,3}=-\frac{{{A_0}C_0^2}}{2}\frac{{(\alpha - {g_1}){\alpha ^2}}}{{({\alpha ^2} + 1)}}\left[ {2{g_1} + \frac{{(\alpha  - {g_1})(2{{k'}^2} + 2{\alpha ^2} - {\alpha ^2}{k^2})}}{{({\alpha ^2} + 1)({\alpha ^2} + {{k'}^2})}}} \right],\quad
g_{2,3}= i c_{2,3}.
\end{array}
\]
Modulus and parameter of elliptic integrals:
\[
k = \frac{{2\sqrt {{A_1}{A_2}}}}{{{A_1} + {A_2}}},\quad
h = {\alpha ^2} + 1.
\]

\begin{acknowledgements}
The work was supported by the Presidium of the Russian Academy of Sciences (Program 28 ``Space: investigations of the fundamental processes and their interrelationships''). We are grateful to A.I.Neishtadt, A.Correia, A.Morbidelli and J.Wisdom for useful discussions. We would also like to thank D.A.Pritykin for proofreading the manuscript.
\end{acknowledgements}

\bibliography{ORbibfile_v3_nodoi}

\end{document}